\documentclass[10pt,aps,prb,twocolumn,superscriptaddress,longbibliography,reprint]{revtex4-2}

\usepackage[utf8]{inputenc}
\usepackage[T1]{fontenc}
\usepackage[english]{babel}
\usepackage{lmodern}

\usepackage{amsmath,amssymb,mathtools,bbm}

\usepackage{graphicx}
\usepackage{tikz}
\usepackage{adjustbox}
\usepackage[shortlabels,inline]{enumitem}
\usepackage{booktabs}

\usepackage[ruled,vlined]{algorithm2e}

\usepackage{silence}
\usepackage[colorlinks=false,allcolors=blue]{hyperref}

\newcommand{\dd}[0]{\mathrm d}
\newcommand{\ee}[0]{\mathrm e}
\newcommand{\ii}[0]{\mathrm i}

\DeclareMathOperator{\tr}{tr}
\DeclarePairedDelimiter{\abs}{\lvert}{\rvert}
\DeclarePairedDelimiter{\norm}{\lVert}{\rVert}
\DeclarePairedDelimiter{\expval}{\langle}{\rangle}
\DeclarePairedDelimiterX{\comm}[2]{[}{]}{#1, #2}
\DeclarePairedDelimiterX{\acomm}[2]{\{}{\}}{#1, #2}
\DeclarePairedDelimiter{\bra}{\langle}{\rvert}
\DeclarePairedDelimiter{\ket}{\lvert}{\rangle}
\DeclarePairedDelimiterX{\braket}[2]{\langle}{\rangle}{#1\!\delimsize\mid\mathopen{}\!#2}
\DeclarePairedDelimiterX{\ketbra}[2]{\vert}{\vert}{#1\,\delimsize\rangle\!\delimsize\langle\,\mathopen{}#2}

\newcommand{\phantk}{{\vphantom{\text{k}}}}
\newcommand{\phantq}{{\vphantom{\text{q}}}}

\newcommand{\syst}{\text{s}}
\newcommand{\systt}{\text{sys}}

\newcommand{\envt}{\text{e}}
\newcommand{\disst}{\text{d}}
\newcommand{\intt}{\text{i}}
\newcommand{\eqt}{\text{eq}}
\newcommand{\advt}{\text{adv}}
\newcommand{\rett}{\text{ret}}
\newcommand{\CLt}{\text{CL}}
\newcommand{\vNt}{\text{vN}}
\newcommand{\fint}{\text{f}}

\newcommand{\timeord}{\mathcal T}
\newcommand{\timeordinv}{\overline{\mathcal T}}

\newcommand{\rarrow}{{\parbox{0.5em}{\tikz{\draw[->](0,0)--(0.5em,0);}}}}
\newcommand{\larrow}{{\parbox{0.5em}{\tikz{\draw[<-](0,0)--(0.5em,0);}}}}

\newcommand{\inlinecode}{\texttt}

\begin{document}

\title{Non-Hermitian Pseudomodes for Strongly Coupled Open Quantum Systems: Unravelings, Correlations and Thermodynamics}

\author{Paul Menczel}
\email{paul@menczel.net}
\affiliation{Theoretical Physics Laboratory, Cluster for Pioneering Research, RIKEN, Wakoshi, Saitama 351-0198, Japan}

\author{Ken Funo}
\affiliation{Institute for Molecular Science, National Institutes of Natural Sciences, Okazaki 444-8585, Japan}

\author{Mauro Cirio}
\affiliation{Graduate School of China Academy of Engineering Physics, Haidian District, Beijing, 100193, China}

\author{Neill Lambert}
\email{nwlambert@gmail.com}
\affiliation{Theoretical Physics Laboratory, Cluster for Pioneering Research, RIKEN, Wakoshi, Saitama 351-0198, Japan}
\affiliation{Quantum Computing Center, RIKEN, Wakoshi, Saitama, 351-0198, Japan}

\author{Franco Nori}
\affiliation{Theoretical Physics Laboratory, Cluster for Pioneering Research, RIKEN, Wakoshi, Saitama 351-0198, Japan}
\affiliation{Quantum Computing Center, RIKEN, Wakoshi, Saitama, 351-0198, Japan}
\affiliation{Physics Department, University of Michigan, Ann Arbor, MI 48109-1040, USA}

\date{\today}

\begin{abstract}
The pseudomode framework provides an exact description of the dynamics of an open quantum system coupled to a non-Markovian environment.
Using this framework, the influence of the environment on the system is studied in an equivalent model, where the open system is coupled to a finite number of unphysical pseudomodes that follow a time-local master equation.
Building on the insight that this master equation does not need to conserve the hermiticity of the pseudomode state, we here ask for the most general conditions on the master equation that guarantee the correct reproduction of the system's original dynamics.
We demonstrate that our generalized approach decreases the number of pseudomodes that are required to model, for example, underdamped environments at finite temperature.
We also provide an unraveling of the master equation into quantum jump trajectories of non-Hermitian states, which further facilitates the utilization of the pseudomode technique for numerical calculations by enabling the use of easily parallelizable Monte Carlo simulations.
Finally, we show that pseudomodes, despite their unphysical nature, provide a natural picture in which physical processes, such as the creation of system-bath correlations or the exchange of heat, can be studied.
Hence, our results pave the way for future investigations of the system-environment interaction leading to a better understanding of open quantum systems far from the Markovian weak-coupling limit.
\end{abstract}

\maketitle

\section{Introduction}

To fully develop modern applications of quantum physics like quantum computers \cite{DowlingPhilosTransRoyalSocA2003, LaddNature2010, BulutaRepProgPhys2011}, quantum simulators \cite{CiracNaturePhys2012, GeorgescuRevModPhys2014} or quantum thermal machines \cite{RossnagelScience2016, JosefssonNatureNanotech2018, PetersonPhysRevLett2019, GuthriePhysRevApplied2022}, it is necessary to understand the behavior of the involved quantum systems accurately.
Such systems unavoidably interact with their macroscopic environments, which are not fully controllable.
Any realistic model must therefore treat these systems as open; that is, it must include the influence of the environment as a stochastic force \cite{Breuer2002, LidarArXiv190200967Quant-Ph2020}.

\begin{figure*}[t]
	\includegraphics[scale=1]{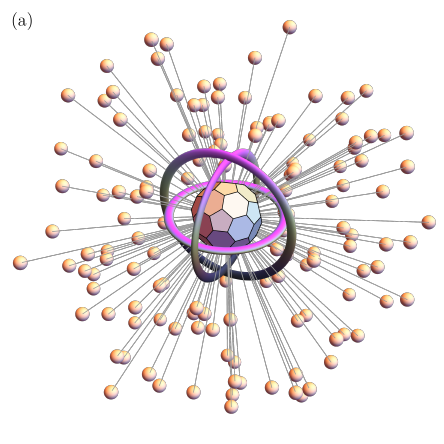}
	\hfill
	\includegraphics[scale=1]{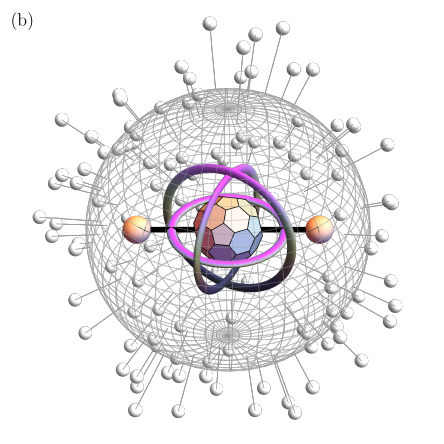}
	\hspace*{\fill}

	\caption{
		Illustration of the pseudomode technique.
		In panel (a), the open system in the middle is directly and strongly coupled to many environment modes.
		Panel (b) depicts an extended system consisting of the open system and two strongly coupled pseudomodes.
		This extended system undergoes a non-unitary and non-positive time evolution, which is represented in the illustration by a weakly coupled residual environment (gray).
		The effects of the environment in panel (a) and the pseudomodes in panel (b) on the open system are equivalent.
	}
	\label{fig:illustration}
\end{figure*}

Until recently, studies of open quantum systems have often focused on the Markovian weak-coupling regime, where it is assumed that the characteristic time scales of system and environment are clearly separated and that their interaction is weak \cite{LidarChemPhys2001, AlbashNewJPhys2012, MajenzPhysRevA2013}.
Under these assumptions, the dynamics of the open quantum system takes the universal shape of a Lindblad master equation \cite{GoriniJMathPhys1976, LindbladCommunMathPhys1976}, which is mathematically well understood and easy to work with both numerically and analytically.

New developments in a variety of research fields (including quantum thermodynamics \cite{Gelbwaser-KlimovskyJPhysChemLett2015, NewmanPhysRevE2017, ThomasPhysRevE2018, AbiusoPhysRevA2019, TalknerRevModPhys2020, MukherjeeCommunPhys2020, DasPhysRevRes2020, CamatiPhysRevA2020, LatunePhysRevAppl2023}, quantum transport \cite{SchallerPhysRevB2018, GehringNatRevPhys2019, PekolaRevModPhys2021, Anto-SztrikacsNewJPhys2021, McConnellNewJPhys2022} and quantum biology \cite{PanitchayangkoonProcNatlAcadSci2011, IshizakiAnnuRevCondensMatterPhys2012, LambertNaturePhys2013, ScholesNature2017}) make it necessary to leave these assumptions behind.
A cornerstone of the theory of open quantum systems beyond the Markovian weak-coupling regime is the Caldeira-Leggett model \cite{CaldeiraPhysicaA1983}.
Despite its conceptual simplicity, this model enjoys a wide applicability and there exist many techniques that can be used to study it; we refer to Ref.~\cite{deVegaRevModPhys2017} for an overview.
This paper focuses on the pseudomode technique \cite{GarrawayPhysRevA1997, TamascelliPhysRevLett2018, LambertNatCommun2019, PleasancePhysRevRes2020, CirioPhysRevRes2023}, where the Caldeira-Leggett model is mapped to a mathematically equivalent Lindblad master equation on an auxiliary Hilbert space.
The auxiliary space comprises the open quantum system and a number of additional ``pseudomodes'', which are calibrated to mimic the behavior of the original Caldeira-Leggett environment, see Fig.~\ref{fig:illustration}.

Even though the pseudomode technique formally uses a Lindblad equation, it involves no approximations and can be applied out of equilibrium, far from the weak-coupling limit, and in the presence of time-dependent driving.
In practice, it is often desirable to use only a (small) finite number of pseudomodes.
The technique then requires the environment auto-correlation function to be written as a finite sum of exponential terms.
This requirement might introduce an approximation, where the true correlation functions are replaced by a finite series expansion or a multi-exponential fit.

We focus on the pseudomode approach due to the following two key advantages.
First, it enables us to transfer tools that have already been developed for Lindblad equations to our strongly coupled setting.
It is therefore comparatively easy to implement in practice.
Second, since environmental degrees of freedom are encoded in the pseudomodes and directly accessible, it provides new insight into the physics of the system-environment interaction.

The method of pseudomodes dates back at least to Garraway's seminal 1997 publication \cite{GarrawayPhysRevA1997}.
However, some key ideas were known in the community already before that, see for example Ref.~\cite{ImamogluPhysRevA1994}.
In Garraway's paper, pseudomodes were used to find the exact dynamics of an atom interacting with a continuum of electromagnetic field modes under the rotating wave approximation.
More recently, the technique was extended to general Caldeira-Leggett models without the rotating wave approximation, going beyond the single-excitation sector \cite{TamascelliPhysRevLett2018}.
It was then generalized to better allow for arbitrary spectral densities by allowing the fitting of generic environment spectral densities \cite{LambertNatCommun2019, MascherpaPhysRevA2020, PleasancePhysRevRes2020}.

The idea of fitting the environment correlation functions or, equivalently, spectral densities with a number of effective modes has recently seen applications in a variety of communities.
Within the field of nanophotonics, for example, this approach is now known as \emph{few-mode field quantization} \cite{MedinaPhysRevLett2021, Sanchez-BarquillaNanophotonics2022, LednevPhysRevLett2024}.
The approach was also applied to fermionic environments \cite{ChenNewJPhys2019, CirioPhysRevRes2023}, where it was demonstrated that it can fully capture the non-trivial physics of the Kondo resonance.
In the context of fermionic environments, it has also been called the \emph{mesoscopic leads} technique \cite{BrenesPhysRevX2020} and applied to thermodynamical questions \cite{LacerdaPhysRevB2023, LacerdaArXiv231212513Quant-Ph2023}.
Another related approach is approximating the continuous environment by a finite number of modes, following unitary closed-system dynamics \cite{BaerJChemPhys1997}, where optimal fitting procedures have recently been discussed in Refs.~\cite{deVegaPhysRevB2015, NusspickelPhysRevB2020}.
Finally, we mention the \emph{quantized quasi-normal mode} technique \cite{FrankePhysRevLett2019} which is enjoying popularity in the field of cavity QED and can be seen as an approximate way of constructing pseudomodes, and we will show in this paper that also the \emph{dissipaton} framework \cite{WangJChemPhys2022, LiArXiv240117255Quant-Ph2024} is closely related to pseudomodes.
There are other approaches that are less directly related, including process tensors \cite{CygorekNatPhys2022, LinkPhysRevLett2024}, the quasi-adiabatic path integral \cite{StrathearnNatCommun2018}, and second-quantized Feshbach projections \cite{LentrodtPhysRevX2020}.

Already in his original publication, Garraway noticed that it may be useful to introduce an effective non-Hermitian Hamiltonian on the atom-pseudomode space by choosing the coupling constants to be complex-valued \cite{GarrawayPhysRevA1997}.
The time evolution on this space then does not keep the full atom-pseudomode state Hermitian, but it still generates the correct physical state of the atom after taking the partial trace over the pseudomode subspace.
The appearance of non-Hermitian pseudomode states thus only highlights their unphysical nature without impacting the usefulness of the method itself.

Recently, in extending the technique to allow for the fitting of generic environments, non-Hermitian couplings have been reintroduced, see for example Refs.~\cite{LambertNatCommun2019, PleasancePhysRevRes2020, CirioPhysRevRes2023, LuoPRXQuantum2023}.
Further, it has been recognized that the introduction of additional complex-valued parameters -- such as complex damping rates \cite{DordaNewJPhys2017} or even temperatures \cite{PleasanceArXiv210805755Quant-Ph2021} -- may allow one to reduce the number of required pseudomodes and thus the computational complexity.
To the authors' knowledge, a systematic analysis of the requirements for the applicability of the pseudomode method is however still missing from the literature, at least for the general case of bosonic pseudomodes involving complex-valued parameters.

Hence, the first goal of this paper is to \emph{clarify the conditions} that the pseudomode Lindblad equation must satisfy to reproduce the system evolution exactly.
To this end, we extend the Feynman-Vernon influence functional formalism, which was originally applied to Caldeira-Leggett environments obeying unitary time evolution \cite{Breuer2002, FeynmanAnnPhysNY1963}, to general master equations on the auxiliary system-pseudomode space.
%This approach makes it possible to precisely identify how the parameters of the pseudomodes must be chosen in order to achieve equivalence with the original Caldeira-Leggett model, and we explain how to perform this matching in practice.
This formalism allows us to establish a rigorous mathematical foundation for the pseudomode framework.
Furthermore, we develop a straightforward recipe that can be used to determine a set of pseudomodes that are equivalent to any given bosonic environment, given a multi-exponential decomposition of the environment's auto-correlation functions.

The second part of this paper aims to \emph{demonstrate the strengths} of this framework and contains two more main results.
The first of these results shows how observables related to the system-environment interaction can be mathematically represented on the auxiliary space.
This result makes it possible to study quantities such as multi-time correlation functions, the system-environment interaction energy or the heat flow from the system to the environments in the pseudomode picture.
This technique is thus able to give insight into the thermodynamics of quantum systems -- a topic that has attracted great interest recently, since its study promises to be useful for a range of topics, from fundamental questions in quantum theory to cutting-edge applications in quantum engineering \cite{PekolaNaturePhys2015, MeraliNature2017}.

For our final main result, we then turn our attention to the study of \emph{quantum jump trajectories}.
In the case of regular Lindblad equations, quantum jump trajectories represent single experimental realizations, somewhat analogous to e.g.\ the stochastic trajectories underlying classical diffusive processes.
They provide access to the full statistics of fluctuating quantities and have thus been studied intensively for a long time, see for example Refs.~\cite{DalibardPhysRevLett1992, DumPhysRevA1992, MolmerJOptSocAmB1993, Carmichael1993, PlenioRevModPhys1998, Breuer2002, BreuerPhysRevA2003, LidarArXiv190200967Quant-Ph2020, MenczelPhysRevRes2020, GneitingPhysRevA2021, GneitingPhysRevRes2022}.
Furthermore, they are a valuable tool for numerical calculations because single trajectories can all be simulated independently in parallel, each at a lower computational cost than that of integrating the full Lindblad equation.
In this paper, we study how the quantum jump trajectory framework can be extended from regular Lindblad equations to the pseudomode framework.
We focus on unravelings of the pseudo-Lindblad equations with complex-valued parameters that can be encountered in the study of pseudomodes, but our discussion in principle also applies to more generalized time-local quantum master equations.

Our results are illustrated using the example of a qubit thermalizing with a strongly coupled thermal environment.
We study its time evolution and the accompanying heat current using pseudomodes and quantum jump trajectories, and benchmark our results against a simulation based on the hierarchical equations of motion (HEOM) \cite{TanimuraJPhysSocJpn1989, KatoThermodynamicsintheQuantumRegime2018, TanimuraJChemPhys2020}.
The HEOM technique can be viewed as formally equivalent to pseudomodes \cite{XuArXiv230716790Quant-Ph2023}, but enjoys a wealth of well-established numerical implementations such as introduced in Ref.~\cite{LambertPhysRevRes2023}.
Furthermore, as a second example, we consider a qubit that is not only coupled to a thermal environment but also subject to an external driving force.
The external driving regularly applies $\pi$-pulses to the system with the effect of reversing the thermalization of the qubit, thus dynamically decoupling it from the environment \cite{ViolaPhysRevA1998}.
The example shows how correlations between qubit and environment behave during this process, and that the pseudomode description can be used to understand them qualitatively.

Our paper is organized as follows.
In Sec.~\ref{sec:pseudo} we introduce our general setup and prove our generalized version of the pseudomode equivalence.
Sec.~\ref{sec:applications} introduces the main results discussed above, which are then illustrated using our examples in Sec.~\ref{sec:examples}.
We conclude in Sec.~\ref{sec:end} and provide perspectives for future research.

\section{The Pseudomode Framework} \label{sec:pseudo}

\subsection{Setup} \label{subsec:cl}

For our general theory, we consider an open quantum system that is coupled to one or more thermal environments.
The pseudomode approach applies to both bosonic and fermionic environments.
To simplify the technical complexity, we only focus only on bosonic environments in this text.
For a discussion of pseudomodes for fermionic environments, we refer to Ref.~\cite{CirioPhysRevRes2023}.

We describe the system and the environments in a generalized Caldeira-Leggett model, where each environment is represented by a continuum of mutually non-interacting harmonic modes \cite{CaldeiraPhysicaA1983, LeggettRevModPhys1987, Breuer2002}.
The Hamiltonian of the total setup therefore has the form
\begin{equation} \label{eq:cl}
	H^\phantq_\CLt(t) = H_\syst^\phantq(t) + \sum\nolimits_\mu \bigl[ H^\mu_\envt + H^\mu_\intt(t) \bigr] .
\end{equation}
We do not make any assumptions about the dimensionality of the system or the shape of the system Hamiltonian $H_\syst(t)$.

The terms $H^\mu_\envt$ and $H^\mu_\intt(t)$ in the total Hamiltonian respectively denote the free Hamiltonian and the interaction Hamiltonian corresponding to the $\mu$-th environment.
They are given by
\begin{equation} \label{eq:cl_Henv}
	H^\mu_\envt \equiv \sum\nolimits_k \omega^\mu_k\, a^{\mu\dagger}_k a^\mu_k ,
\end{equation}
where $\omega^\mu_k$ is the frequency of the $k$-th mode of this environment and $a^\mu_k$ the corresponding annihilation operator, and
\begin{equation}
	H^\mu_\intt(t) \equiv Q^\mu_\phantk(t) X^\mu_\phantk ,
\end{equation}
where $Q^\mu(t)$ is a dimensionless coupling operator on the system Hilbert space and $X^\mu \equiv \sum_k g^\mu_k / \sqrt{2\omega^\mu_k} (a^\mu_k + a^{\mu\dagger}_k)$ the bath coordinate.
We note that, in principle, our results are not restricted to environment Hamiltonians of this exact form as long as certain basic Gaussianity assumptions are satisfied, see Sec.~\ref{subsec:pm}.

The coupling coefficients $g^\mu_k$ are typically specified in the form of a spectral density
\begin{equation}
	G^\mu(\omega) \equiv \pi \sum\nolimits_k \frac{(g^\mu_k)^2}{2\omega^\mu_k}\, \delta(\omega - \omega^\mu_k) .
\end{equation}
%with $G^\mu(\omega) = 0$ for $\omega \leq 0$.
Note that we set $\hbar$ to one throughout this text, and that we allow the system Hamiltonian and the system coupling operators to explicitly depend on the time $t$ in order to model time-dependent driving.
We denote the state of the total setup by $\rho_\CLt(t)$; it follows the unitary time evolution
\begin{equation}
	\partial_t \rho_\CLt(t) = -\ii \comm{H_\CLt(t)}{\rho_\CLt(t)} .
\end{equation}

A central quantity characterizing the environments is their free two-time correlation functions $C^\mu(t)$.
We make the assumption that the state of the system and environments factorizes at the initial time $t=0$, and that each environment starts in a canonical equilibrium state
\begin{equation} \label{eq:thermal_initial_state}
	\rho^\mu_\eqt \propto \exp[ -\beta_\phantq^\mu H^\mu_\envt ] .
\end{equation}
The case of non-equilibrium initial states will also be discussed later.
Here, $\beta^\mu$ denotes the inverse temperature of the environment and the proportionality factor is fixed by the normalization of the state.
A short calculation shows that under these assumptions, the correlation functions are given by
\begin{align} \label{eq:cl_cfct}
	C^\mu_\phantq(t) &\equiv \tr[ X_\phantq^\mu(t) X_\phantq^\mu \rho^\mu_\eqt ] \nonumber \\
		&= \int_{-\infty}^\infty\! \frac{\dd\omega}{\pi} G^\mu(\omega) \Bigl[ \coth\Bigl( \frac{\beta^\mu\omega}{2} \Bigr) \cos(\omega t) - \ii \sin(\omega t) \Bigr] ,
\end{align}
where $X^\mu(t) \equiv \ee^{\ii H^\mu_\envt t}\, X^\mu\, \ee^{-\ii H^\mu_\envt t}$ denotes the Heisenberg picture operator.
For later reference, we note that $C^\mu(-t) = C^\mu(t)^\ast$ is the complex conjugate and that
\begin{equation} \label{eq:langevin_question_mark}
	\comm{X^\mu(t)}{X^\mu(t')} = 2\ii\, \Im[ C^\mu(t - t') ]
\end{equation}
holds for any two times $t$ and $t'$, with $\Im$ being the imaginary part.
This equation follows immediately from the fact that the commutator of $X^\mu(t)$ and $X^\mu(t')$ is always a scalar.

The Fourier transform of the correlation functions,
\begin{equation}
	\int_{-\infty}^\infty \dd t\, C^\mu(t)\, \ee^{\ii \omega t} = \bigl[ G^\mu(\omega) - G^\mu(-\omega) \bigr] \Bigl[ 1 + \coth\Bigl( \frac{\beta^\mu \omega} 2 \Bigr) \Bigr] ,
\end{equation}
reveals that they contain combined information about the bath temperatures and spectral densities.
Neither of these quantities can be independently recovered from the correlation functions.
Nevertheless, the specification of these correlation functions alone already fully determines the dynamics of the open system \cite{Breuer2002}.
That is, aside from the system operators $H_\syst(t)$ and $Q^\mu(t)$, the reduced state
\begin{equation}
	\rho_\syst(t) \equiv \tr_\envt \rho_\CLt(t)
\end{equation}
of the system depends only on the functions $C^\mu(t)$.
Here, $\tr_\envt$ denotes the partial trace over all environments.

Different environments with identical free correlation functions are thus equivalent from the open system's point of view.
The core idea of the pseudomode method is to make use of this equivalence to \emph{replace the original environment with one that can be treated more easily}.
To maximize the freedom in choosing the replacement, the equivalence is extended beyond the unitary time evolution discussed so far, as we will see in the next section.

\subsection{Non-Unitary Environments} \label{subsec:pm}

Let us now consider an auxiliary Hilbert space consisting of the open system and the replacement environment, and a state $\rho(t)$ that obeys a non-unitary time evolution equation.
We will here assume that it has the form
\begin{align}
	\partial_t \rho(t) = &-\ii \comm[\big]{H_\syst(t)}{\rho(t)} \nonumber \\
		&+ \sum\nolimits_{\mu n} \Bigl( \mathbf L^\mu_n\, \rho(t) - \ii \lambda^\mu_n\, \comm[\big]{Q^\mu_\phantq(t) X^\mu_n}{\rho(t)} \Bigr) \label{eq:pm_evo}
\end{align}
and discuss further generalizations -- such as a Tanimura terminator, which is an extra non-unitary term acting on the system subspace -- in Appendix \ref{app:influence_generalizations}.
For notational convenience, we have divided the replacement environment into one or more auxiliary subspaces for each environment of the original model.
These auxiliary subspaces will later correspond to the pseudomodes.
The ranges of their indices $n$ may differ depending on the associated environment.
The linear superoperators $\mathbf L^\mu_n$ which describe the free evolution and the coupling operators $X^\mu_n$ only act on their respective subspaces, and $\lambda^\mu_n$ are coupling constants.
The Caldeira-Leggett model described previously is a special case of this form with only one auxiliary subspace per environment, $\mathbf L^\mu\, \bullet = -\ii \comm{H^\mu_\envt}{\bullet}$ and $\lambda^\mu = 1$.

Our first main result is that within the class of models described by Eq.~\eqref{eq:pm_evo}, given three basic assumptions described below, different environments are still equivalent as long as their correlation functions are identical.
The correlation functions are
\begin{align}
	C_\advt^\mu(t) &\equiv \sum\nolimits_n (\lambda^\mu_n)^2 \tr\bigl[ X^\mu_n(t) X^\mu_n\, \rho^\mu_{\eqt,n} \bigr] \quad \text{and} \nonumber\\
	C_\rett^\mu(t) &\equiv \sum\nolimits_n (\lambda^\mu_n)^2 \tr\bigl[ X^\mu_n X^\mu_n(t)\, \rho^\mu_{\eqt,n} \bigr] , \label{eq:cfct_def}
\end{align}
both evaluated at times $t \geq 0$.
In this definition,
\begin{equation} \label{eq:heisenberg}
	X^\mu_n(t) \equiv (\ee^{\mathbf L^\mu_n t})^\dagger X^\mu_n
\end{equation}
are the coupling operators in the Heisenberg picture, the dagger signifies the adjoint with respect to the Hilbert-Schmidt inner product, and $\rho^\mu_{\eqt,n}$ denotes the stationary state of the generator $\mathbf L^\mu_n$.

More generally, the stationary state could be replaced by any Gaussian initial state $\rho^\mu_{0,n}$.
In this case, the correlation functions depend explicitly on both times, instead of just on the time difference:
\begin{align}
	C_\advt^\mu(t, t') &\equiv \sum\nolimits_n (\lambda^\mu_n)^2 \tr\bigl[ X^\mu_n(t-t') X^\mu_n\, \rho^\mu_{0,n}(t') \bigr] \quad \text{and} \nonumber\\
	C_\rett^\mu(t, t') &\equiv \sum\nolimits_n (\lambda^\mu_n)^2 \tr\bigl[ X^\mu_n X^\mu_n(t-t')\, \rho^\mu_{0,n}(t') \bigr] ,
\end{align}
where $\rho^\mu_{0,n}(t') = \exp[ \mathbf L^\mu_n t'] \rho^\mu_{0,n}$.
More details on non-stationary initial states can be found in Appendix~\ref{app:influence_generalizations}.

For unitary environments, the advanced correlation function $C_\advt^\mu(t)$ agrees with the correlation function as defined in Eq.~\eqref{eq:cl_cfct}, and the retarded correlation function $C_\rett^\mu(t)$ is its complex conjugate.
In the literature, the two correlation functions are therefore sometimes bundled into a single object $C^\mu(t)$ which is defined as $C_\advt^\mu(t)$ for $t \geq 0$ and as $C_\rett^\mu(\abs{t})$ for $t \leq 0$.
However, this alternative definition can lead to confusion since, for non-unitary dynamics,
\begin{equation}
	\tr\bigl[ X^\mu_n(-t) X^\mu_n\, \rho^\mu_{\eqt,n} \bigr] \neq \tr\bigl[ X^\mu_n X^\mu_n(t)\, \rho^\mu_{\eqt,n} \bigr]
\end{equation}
in general.
We prefer to treat $C_\advt^\mu(t)$ and $C_\rett^\mu(t)$ as two separate functions, both defined only at $t \geq 0$, and consider them to be independent degrees of freedom.
Hence, to gain access to the larger class of environments, we must pay the price of matching two correlation functions instead of one.
We shall see that this undertaking is beneficial nevertheless.

We now return to the three basic assumptions underlying our result, which are as follows:

\begin{enumerate*}[(i)]
\item
	The generators $\mathbf L^\mu_n$ must be trace-\allowbreak preserving, that is, $\tr[ \mathbf L^\mu_n \rho ] = 0$.\\

\item
	They must have Gaussian stationary states $\rho^\mu_{\eqt,n}$, i.e., the moments of the stationary states must obey Wick's probability theorem.
	A precise formulation of this requirement can be found in Eq.~\eqref{eq:app:wick_theorem} in the Appendix.
	In the previously mentioned case of non-stationary initial states, the initial states must still be Gaussian.\\

\item
	In order to state the third assumption, we define, for any operator, $A$ the superoperators
\end{enumerate*} % Can't have display math inside enumerate* but it works like this
	\begin{align}
		(A)^\larrow_t\, \bullet &\equiv \ee^{-\mathbf L_\envt t} \bigl[ (\ee^{\mathbf L_\envt t} \bullet) A \bigr] \text{ and} \nonumber\\
		(A)^\rarrow_t\, \bullet &\equiv \ee^{-\mathbf L_\envt t} \bigl[ A (\ee^{\mathbf L_\envt t} \bullet) \bigr] \label{eq:supop_interaction_pic}
	\end{align}
	with $\mathbf L_\envt = \sum_{\mu n} \mathbf L^\mu_n$.
	The third assumption is that the commutator of $(X^\mu_n)^i_t$ with $(X^\mu_n)^j_{t'}$ is a complex number for any two times $t$ and $t'$ and for any combination of $i, j \in \{ \larrow, \rarrow \}$.
	That is, we require that
	\begin{equation} \label{eq:basic_assumption3}
		\comm{(X^\mu_n)^i_t}{(X^\mu_n)^j_{t'}} \propto \mathbf 1 ,
	\end{equation}
	where $\mathbf 1$ is the identity superoperator.

Assumption (iii) can be viewed as a generalization of the relation \eqref{eq:langevin_question_mark}, which holds in unitary environments and implies that commutators of bath coordinate operators in the Heisenberg picture are complex numbers.
Intuitively, this assumption is also related to Gaussianity, since a Gaussianity-preserving time evolution also preserves the linearity of the coupling operators, and maps the bath coordinate operator to a linear combination of the bath coordinate and momentum operators.

While assumptions (i) to (iii) are strong constraints, we point out that they still leave much freedom.
The coupling constants $\lambda^\mu_n$ do not need to be real; they can be arbitrary complex numbers as in Refs.~\cite{LambertNatCommun2019, PleasancePhysRevRes2020, CirioPhysRevRes2023}.
The superoperators $\mathbf L^\mu_n$ are not required to generate positive maps and they may even violate $(\mathbf L^\mu_n \rho)^\dagger = \mathbf L^\mu_n (\rho^\dagger)$.
Finally, the stationary states $\rho^\mu_{\eqt,n}$ are, unlike in Refs.~\cite{LambertNatCommun2019, PleasancePhysRevRes2020, CirioPhysRevRes2023}, not required to be Hermitian.
We are thus considering potentially very unphysical time evolutions on the replacement environments.
However, as long as these environments are equivalent to the original unitary environment, the resulting reduced dynamics of the open system will be the same as the original one, and therefore physical.

A crucial ingredient in our main result is that the original Caldeira-Leggett environment itself satisfies our three assumptions, and can therefore be substituted with a different environment.
We note that this result can still be applied if the original environment is not of the form \eqref{eq:cl_Henv}, as long as the three assumptions hold.
However, in the form stated here, the result does not immediately extend to fermionic environments.
In fermionic environments, the commutation relation \eqref{eq:basic_assumption3} typically does not hold and would be replaced with an anti-commutation relation; also, Wick's theorem does not hold in the form given in Eq.~\eqref{eq:app:wick_theorem}, which ignores signs picked up from anti-commuting fermionic operators.
As stated in the beginning, this paper thus focuses on bosonic environments only, and we refer to Ref.~\cite{CirioPhysRevRes2023} for an explanation of how fermions can be accommodated.

We finally move to the proof of our main result. In Appendix \ref{app:influence_functional}, we show that the reduced state of the system can be written in terms of a Feynman-Vernon influence functional:
\begin{equation} \label{eq:influence_functional}
	\rho_\syst(t) = \timeord\Bigl[ \ee^{-\ii \int_0^t \dd\tau\, H_\syst(\tau)^\times} \Bigr]\, \timeord\Bigl[ \ee^{\sum_\mu \int_0^t \dd\tau\, \mathbf W^\mu(\tau)} \Bigr]\, \rho_\syst(0) ,
\end{equation}
where $\timeord$ denotes time ordering, with later times moved to the left.
The influence phase superoperators $\mathbf W^\mu(\tau)$ are given by
\begin{align}
	\mathbf W^\mu_\phantk(\tau) \equiv &- \int_0^\tau \dd\tau'\, C_\advt^\mu(\tau - \tau')\, \tilde Q_\phantk^\mu(\tau)^\times \tilde Q_\phantk^\mu(\tau')^\rarrow \nonumber\\
		&+ \int_0^\tau \dd\tau'\, C_\rett^\mu(\tau-\tau')\, \tilde Q_\phantk^\mu(\tau)^\times \tilde Q_\phantk^\mu(\tau')^\larrow . \label{eq:influence_phase}
\end{align}
Here, we defined for any operator $A$ the superoperators $A^\larrow$ and $A^\rarrow$ acting as $A^\larrow\, \bullet \equiv \bullet\, A$ and $A^\rarrow\, \bullet \equiv A\, \bullet$, as well as $A^\times \equiv A^\rarrow - A^\larrow$.
The notation $\tilde Q^\mu(t)$ indicates that the operator is expressed in the interaction picture with respect to $H_\syst(t)$.
We thus find that the reduced state only depends on the two correlation functions defined in Eq.~\eqref{eq:cfct_def}, as claimed.

\subsection{Pseudomodes} \label{subsec:harmonic_pms}

\begin{table*}[t]
	\centering
	\begin{tabular}{l@{\extracolsep{1cm}}lll@{\extracolsep{.5cm}}lll}
		\toprule
		\textbf{CF Term} & \textbf{Conditions} & \textbf{PMs} & \multicolumn{4}{l}{\textbf{Pseudomode parameters}} \\
		&&& $\Omega$ & $\Gamma$ & $N$ & $\lambda^2$ \\ \midrule
		$a \ee^{-\nu t}$ & $a \in \mathbb R$ & $1$ & $\Im(\nu)$ & $2\Re(\nu)$ & $0$ & $a$ \\ \midrule
		$a \ee^{-\nu_1 t} + a^\ast \ee^{-\nu_2 t}$ & - & $2$ & $\frac{1}{2\ii} (\nu_1 - \nu_2^\ast)$ & $\nu_1 + \nu_2^\ast$ & $0$ & $a$ \\ \addlinespace[0.2em]
		&&& $\frac{1}{2\ii} (\nu_2 - \nu_1^\ast)$ & $\nu_2 + \nu_1^\ast$ & $0$ & $a^\ast$ \\ \addlinespace[0.2em] \midrule
		$a \ee^{-\nu_1 t} - a \ee^{-\nu_2 t}$ & - & $3$ & $\Im(\nu_1)$ & $2\Re(\nu_1)$ & $0$ & $a^\ast$ \\
		&&& $\Im(\nu_2)$ & $2\Re(\nu_2)$ & $0$ & $-a$ \\
		&&& $\frac{1}{2\ii} (\nu_1 - \nu_2^\ast)$ & $\nu_1 + \nu_2^\ast$ & $0$ & $a - a^\ast$ \\ \addlinespace[0.2em] \midrule
		$a_1 \ee^{-\nu t} + a_2 \ee^{-\nu^\ast t}$ & $a_1 + a_2 \in \mathbb R$ & $2$ & $\Im(\nu)$ & $2\Re(\nu)$ & $a_2 / (a_1 - a_2^\ast)$ & $a_1 - a_2^\ast$ \\
		& and $\abs{\Re(a_1)} > \abs{\Re(a_2)}$ && $0$ & $2\nu$ & $0$ & $a_1 - a_1^\ast$ \\ \midrule
		$a \ee^{-\nu t}$ & $a \notin \mathbb R$ & \multicolumn{5}{l}{Treat as $a \ee^{-\nu t} + a^\ast \ee^{-\Omega t}$ for $\Omega \to \infty$.} \\ \bottomrule
	\end{tabular}
	\caption{
		Constructing pseudomodes (PMs) for a given environment correlation function (CF).
		We assume that the CF is a sum of terms of the types listed in the first column.
		The second column gives restrictions on the parameters appearing in the respective CF term.
		Parameters not mentioned here may be arbitrary complex numbers.
		The third column lists the number of PMs required to match the respective term, and the remaining columns list the PM parameters as functions of the CF parameters.
		Each PM is fully specified by the four parameters $\Omega^\mu_n$, $\Gamma^\mu_n$, $N^\mu_n$ and $\lambda^\mu_n$, see Eqs.~\eqref{eq:pm_evo} and \eqref{eq:pm_lindblad}.
		The symbols $\Re$ and $\Im$ denote the real and imaginary parts of complex numbers.
		The last line of this table describes the regularization procedure mentioned in the main text.
	}
	\label{tab}
\end{table*}

To construct a concrete replacement for a given unitary environment with the correlation function $C^\mu(t)$, we need to find one with matching correlation functions $C_\advt^\mu(t)$ and $C_\rett^\mu(t)$, that is,
\begin{equation}
	C_\advt^\mu(t) = C_\phantk^\mu(t)
	\quad \text{and} \quad
	C_\rett^\mu(t) = C_\phantk^\mu(t)^\ast .
\end{equation}
Here, the asterisk denotes complex conjugation.
In order to proceed, we must now specify the auxiliary subspaces, which so far have been kept fully general.
We will focus on the dynamics
\begin{align}
	\mathbf L^\mu_n \rho &= -\ii \Omega^\mu_n\, \comm{ b_n^{\mu \dagger} b^\mu_n }{\rho} \nonumber \\
		&\quad + \Gamma^\mu_n (N^\mu_n {+} 1)\, \bigl( b^\mu_n\, \rho\, b_n^{\mu \dagger} - \acomm{b_n^{\mu \dagger} b^\mu_n}{\rho} / 2 \bigr) \nonumber \\
		&\quad + \Gamma^\mu_n N^\mu_n\, \bigl( b_n^{\mu \dagger} \rho\, b^\mu_n - \acomm{b^\mu_n b_n^{\mu \dagger}}{\rho} / 2 \bigr) , \label{eq:pm_lindblad}
\end{align}
where $b_n^\mu$ and $b_n^{\mu \dagger}$ are bosonic ladder operators and the curly braces denote the anti-commutator, and choose
\begin{equation}
	X^\mu_n = b^\mu_n + b^{\mu\dagger}_n .
\end{equation}
The symbols of the free parameters $\Omega^\mu_n$, $\Gamma^\mu_n$ and $N^\mu_n$ have been chosen to make this generator resemble the standard Lindbladian for a damped harmonic oscillator.
However, both $\Omega^\mu_n$ and $\Gamma^\mu_n$ can be arbitrary complex values, and $N^\mu_n$ a complex value with a real part greater than $-1/2$.
To highlight this fact, we call these unphysical modes pseudomodes and Eq.~\eqref{eq:pm_evo} a pseudo-Lindblad equation \cite{LambertNatCommun2019, LambertArXiv231012539Quant-Ph2023}.

These pseudomodes are the most straightforward implementation of an environment satisfying the three assumptions of Sec.~\ref{subsec:pm}.
To further generalize the approach, it would be possible to consider additional terms in the time evolution that preserve Gaussianity.
One could thus add terms to the pseudomode Hamiltonian that are linear or quadratic in the pseudomode coupling operators, and one could even consider multiple coupled pseudomodes with bilinear coupling terms.
Coupled pseudomodes have been discussed, for example, in Refs.~\cite{MascherpaPhysRevA2020, MedinaPhysRevLett2021, Sanchez-BarquillaNanophotonics2022, LentrodtPhysRevLett2023, LednevPhysRevLett2024}.
Our framework also encompasses dissipatons \cite{WangJChemPhys2022, LiArXiv240117255Quant-Ph2024}, which formally look like pseudomodes in pure states and may therefore be more efficient numerically, see Appendix~\ref{app:dissipatons}.

In the following, we will however continue to focus on the simple pseudo\-modes \eqref{eq:pm_lindblad}, because they provide the clearest physical picture and are sufficiently flexible for our applications without too much technical complication.

The stationary state of the generator \eqref{eq:pm_lindblad} is given by
\begin{equation} \label{eq:lindblad_ss}
	\rho^\mu_{\eqt,n} \propto \exp\bigl[ -\log[ (N^\mu_n {+} 1) / N^\mu_n ]\, b_n^{\mu \dagger} b^\mu_n \bigr] ,
\end{equation}
where the proportionality factor is fixed by the normalization $\tr \rho^\mu_{\eqt,n} = 1$.
Note that the state is normalizable because of the real part of $N^\mu_n$ being greater than $-1/2$.

In Appendix \ref{app:pms}, we show that the generator satisfies all three assumptions of Sec.~\ref{subsec:pm}.
We also show that with this dynamics, the contribution of a single pseudomode to the correlation functions \eqref{eq:cfct_def} evaluates to
\begin{align}
	&\tr\bigl[ X^\mu_n(t) X^\mu_n\, \rho^\mu_{\eqt,n} \bigr] \nonumber\\
		&\quad = N^\mu_n\, \ee^{\ii \Omega^\mu_n t - \Gamma^\mu_n t / 2} + (N^\mu_n {+} 1)\, \ee^{-\ii \Omega^\mu_n t - \Gamma^\mu_n t / 2} \quad \text{and} \nonumber\\
	&\tr\bigl[ X^\mu_n X^\mu_n(t)\, \rho^\mu_{\eqt,n} \bigr] \nonumber\\
		&\quad = N^\mu_n\, \ee^{-\ii \Omega^\mu_n t - \Gamma^\mu_n t / 2} + (N^\mu_n {+} 1)\, \ee^{\ii \Omega^\mu_n t - \Gamma^\mu_n t / 2} . \label{eq:pms_cfcts}
\end{align}

This result makes it clear that a finite number of pseudomodes can only exactly match a correlation function that is a finite sum of exponential terms.
However, the spectral densities describing the environments are often derived phenomenologically and the correlation functions in that case unavoidably come with some uncertainty \cite{MascherpaPhysRevLett2017}.
It is therefore often justified to apply multi-exponential approximations to the correlation functions.
Furthermore, we physically expect that a slight change in the environment correlation functions changes the dynamics of the open system only slightly {\cite{MascherpaPhysRevLett2017, TrivediPhysRevLett2021}}.
Multi-exponential expansions can thus still provide good approximations even when the environment spectral densities are known exactly from first principles.
In practice, multi-exponential expansions of the correlation functions can be obtained from pole expansions of Eq.~\eqref{eq:cl_cfct}, from numerical fitting procedures \cite{LambertNatCommun2019}, through heuristic approaches like in Ref.~\cite{BrenesPhysRevX2020}, by Prony analysis \cite{WangJChemPhys2022, LiArXiv240117255Quant-Ph2024}, or from rational approximations of the spectral density \cite{NakatsukasaSIAMJSciComput2018, XuPhysRevLett2022}.
We refer to Ref.~\cite{TakahashiJChemPhys2024} for a recent review of these and other approaches.

In Table~\ref{tab}, we provide a dictionary mapping possible terms in a multi-exponential correlation function to one or more pseudomodes mimicking these terms.
Note that the shape of the terms considered here stems from the fact that $C^\mu_\advt(0) = C^\mu_\rett(0)$ holds in general; therefore, only correlation functions with $C^\mu(0) \in \mathbb R$ can be handled.
For environments violating this property, such as overdamped Drude-Lorentz environments, one must add a regularization term as indicated at the end of the table.
We will demonstrate this regularization procedure later in an example, see Sec.~\ref{subsec:ex2}.

\subsection{Comparison with Reaction Coordinates}
We note that the equivalent pseudomode model that we have constructed here resembles the reaction coordinate model \cite{Iles-SmithPhysRevA2014, StrasbergNewJPhys2016, StrasbergPhysRevB2018} where, similarly, a number of environmental degrees of freedom are treated exactly.
The reaction coordinate approach is based on a transformation of the environment which singles out the reaction coordinates, i.e., the most strongly interacting degrees of freedom.
After extracting sufficiently many reaction coordinates, the residual bath can be coupled more weakly and is often treated in a Markovian weak-coupling approximation.
Recent extensions of the reaction coordinate technique include effective modes \cite{HughesJChemPhys2009}, chain mappings \cite{ChinJMathPhys2010} and the reaction coordinate polaron transform \cite{Anto-SztrikacsPRXQuantum2023}.

The two techniques differ in the following two points.
First, the pseudomode technique provides a direct mapping of the unitary environment to a master equation for the system and the pseudomodes.
This mapping is exact to the degree that the environment correlation functions can be fitted with, or otherwise approximated by, a multi-exponential function.
To the contrary, the reaction coordinate approach only maps one Caldeira-Leggett type model to another, and the environment in the new model must still be dealt with, for example through a Markovian approximation.
Second, reaction coordinates are physical modes and thus have a transparent interpretation in contrast to pseudomodes.
In this paper, we have so far introduced pseudomodes as only a mathematical tool.
Their properties are related to the abstract exponents and coefficient of the multi-exponential correlation function, and they are in principle unphysical.
However, we will see in the following that they can often still be treated as if they were physical degrees of freedom representing the environment.

\section{Applications} \label{sec:applications}

\subsection{Multi-Time Correlation Functions}

Our considerations so far allow us to find the reduced state of the open quantum system.
We are thus able to compute expectation values of system variables, averaging over the environmental degrees of freedom.
However, in order to study response properties of the system or fluctuations of its dynamical variables, we must go a step further.
The quantities of interest are then multi-time correlation functions \cite{AlonsoPhysRevLett2005}, which have the general form
\begin{align}
	f &= \tr\bigr[ S_k \mathbf V_\CLt(t_k, t_{k-1}) \cdots S_1 \mathbf V_\CLt(t_1, 0) \rho_\CLt(0) \bigl] \nonumber\\
	  &= \tr\bigr[ S_k(t_k) \cdots S_1(t_1) \rho_\CLt(0) \bigl] . \label{eq:multi_time_cfct_CL}
\end{align}
Here, the $S_i$ are system operators, and the times $t_i$ are ordered such that $t_k \geq \cdots \geq t_1 \geq 0$.
The propagation superoperators $\mathbf V_\CLt(t_b, t_a) \equiv \timeord \exp\bigl[ -\ii \int_{t_a}^{t_b} \dd t\, H_\CLt(t)^\times \bigr]$ act on everything to their right, and $S_i(t_i) \equiv \mathbf V_\CLt(t_i, 0)^\dagger S_i$ denotes Heisenberg picture operators.

The first expression in Eq.~\eqref{eq:multi_time_cfct_CL} readily generalizes to non-unitary environments, where we define
\begin{align}
	f &\equiv \tr\bigr[ S_k \mathbf V(t_k, t_{k-1}) \cdots S_1 \mathbf V(t_1, 0) \rho(0) \bigl] \nonumber\\
	  &= \tr\bigr[ (S_k)^\rarrow_{t_k} \cdots (S_1)^\rarrow_{t_1} \rho(0) \bigl] . \label{eq:multi_time_cfct}
\end{align}
Here, $\mathbf V(t_b, t_a) \equiv \timeord \exp\bigl[ \int_{t_a}^{t_b} \dd t\, \mathbf L_0(t) \bigr]$ is the propagator for the time evolution equation \eqref{eq:pm_evo}, written as $\partial_t \rho(t) = \mathbf L_0(t) \rho(t)$, and the superoperators $(S_i)^\rarrow_{t_i}$ are defined as $(S_i)^\rarrow_{t_i} \equiv \mathbf V(t_i, 0)^{-1} S_i^\rarrow \mathbf V(t_i, 0)$.

For a unitary environment, the new definition \eqref{eq:multi_time_cfct} obviously reduces to Eq.~\eqref{eq:multi_time_cfct_CL}.
However, it is not \emph{a priori} clear whether this expression produces the same value with any equivalent environment.
In the remainder of this section, we shall prove that it does.
In other words, we will show that the expression depends only on system space operators and the correlation functions $C^\mu_\advt(t)$ and $C^\mu_\rett(t)$.
We note that this equivalence has already been shown in Ref.~\cite{SmirneOpenSystInfDyn2022} for pseudomodes with physical parameters.
We present a different proof here that is more general and less technical.

For our proof, we take inspiration from a technique that was used in Ref.~\cite{KatoJChemPhys2016} to compute certain expectation values in the hierarchical equations of motion framework.
We add source fields $J_i(t)$ to the time evolution equation \eqref{eq:pm_evo}, modifying it into
\begin{equation}
	\partial_t \rho_J(t) = \mathbf L_0(t)\, \rho_J(t) - \ii \sum\nolimits_i J_i(t) S_i\, \rho_J(t)
\end{equation}
with the formal solution
\begin{equation}
	\rho_J(t) = \timeord \exp\biggl\{ \int_0^t \dd\tau \left[ \mathbf L_0(\tau) - \ii \sum\nolimits_i J_i(\tau) S_i^\rarrow \right] \biggr\}\, \rho(0) .
\end{equation}
The formal solution shows that the multi-time correlation function in question can be obtained by taking functional derivatives of the time-ordered exponential,
\begin{equation}
	f = \frac{1}{(-\ii)^k} \tr\Bigl[ \frac{\delta}{\delta J_1(t_1)} \cdots \frac{\delta}{\delta J_k(t_k)} \tr_\envt \rho_J(t) \Bigr]_{J=0} .
\end{equation}

It remains to show that the time evolution of $\tr_\envt \rho_J(t)$ only depends on system operators and the bath correlation functions.
If this invariance holds for any source fields, it will also hold for the functional derivatives in the equation above.
Repeating the derivation of the influence functional with the modified time evolution, we find
\begin{equation} \label{eq:mtcf_proof}
	\tr_\envt \rho_J(t) = \timeord\Bigl[ \ee^{-\ii \int_0^t \dd\tau\, H_\syst(\tau)^\times} \Bigr]\, \timeord\Bigl[ \ee^{\int_0^t \dd\tau\, \mathbf W_J(\tau)} \Bigr]\, \rho_\syst(0) ,
\end{equation}
see Appendix \ref{app:influence_functional}.
The result is almost identical to Eq.~\eqref{eq:influence_functional}, except that the source fields have modified the total influence phase superoperator into
\begin{equation}
	\mathbf W_J(\tau) = \sum\nolimits_\mu \mathbf W^\mu(\tau) - \ii \sum\nolimits_i J_i(\tau) \tilde S_i(\tau)^\rarrow ,
\end{equation}
with $\mathbf W^\mu(\tau)$ as defined in Eq.~\eqref{eq:influence_phase}.
The interaction picture operators $\tilde S_i(\tau)$ depend only on system operators.
Hence, the whole expression \eqref{eq:mtcf_proof} depends only on system space operators and on the bath correlation functions, and our proof is complete.

The multi-time correlation functions \eqref{eq:multi_time_cfct_CL} can therefore be easily computed in pseudomode models using Eq.~\eqref{eq:multi_time_cfct}.
Moreover, our proof could be easily generalized to thermal correlation functions, where the initial product state $\rho_\CLt(0)$ is replaced by the combined system-environment equilibrium state $\rho_{\CLt,\eqt}$.
To do so, one would assume that the equilibrium state can be written as
\begin{equation} \label{eq:thermal_corr_fct}
	\rho_{\CLt,\eqt} = \mathbf V_\CLt(0, -T) \rho_\CLt(-T)
\end{equation}
for $T \to \infty$ and some product state $\rho_\CLt(-T)$, and then proceed as described above.
This generalization was discussed in Ref.~\cite{CirioPhysRevB2022} for two-time correlation functions, but it can be done in our formalism in a more straightforward manner.

\subsection{System-Environment Interactions}

To study the interaction between an open quantum system and its environment, it is often necessary to calculate expectation values involving one or more copies of interaction Hamiltonians.
Since the bath coupling operators commute with each other and with operators on the system space, any such expectation value can be brought into the generic form
\begin{equation} \label{eq:expectation_values_cl}
	E = \tr\bigl[ S\, X^{\mu_1} \cdots X^{\mu_k}\, \rho_\CLt(t) \bigr] .
\end{equation}
Here, $S$ is a system operator and $X^{\mu_1} \cdots X^{\mu_k}$ a string of $k$ (not necessarily distinct) environment coupling operators.

Our goal is to compute the expectation value $E$ in a pseudomode model.
To this end, we proceed like in the previous section.
That is, we first generalize the expression \eqref{eq:expectation_values_cl} to non-unitary environments,
\begin{equation} \label{eq:expectation_values}
	E \equiv \tr\Bigl[ S\, \Bigl( \sum\nolimits_n \lambda^{\mu_1}_n X^{\mu_1}_n \Bigl)\, \cdots \Bigl( \sum\nolimits_n \lambda^{\mu_k}_n X^{\mu_k}_n \Bigl)\, \rho(t) \Bigl] ,
\end{equation}
and then show that the new expression only depends on system operators and the bath correlation functions.
Note that the indices $n$ may range over different values in the different sums, depending on the number of pseudomodes that are used to represent the associated environment.

Adding again source fields to the time evolution equation \eqref{eq:pm_evo}, it now becomes
\begin{equation} \label{eq:source_field}
	\partial_t \rho_J(t) = \mathbf L_0(t)\, \rho_J(t) - \ii \sum\nolimits_\mu J^\mu(t) \Bigl( \sum\nolimits_n \lambda^\mu_n X^\mu_n \Bigr) \rho_J(t) .
\end{equation}
The expectation value in question can be obtained by taking functional derivatives of the time-ordered exponential,
\begin{equation}
	E = \frac{1}{(-\ii)^k} \tr\Bigl[ S\, \frac{\delta}{\delta J^{\mu_1}(t)} \cdots \frac{\delta}{\delta J^{\mu_k}(t)} \tr_\envt \rho_J(t) \Bigr]_{J=0} .
\end{equation}
The influence functional representation of $\tr_\envt \rho_J(t)$ is
\begin{equation} \label{eq:interaction_proof}
	\tr_\envt \rho_J(t) = \timeord\Bigl[ \ee^{-\ii \int_0^t \dd\tau\, H_\syst(\tau)^\times} \Bigr]\, \timeord\Bigl[ \ee^{\sum_\mu \int_0^t \dd\tau\, \mathbf W^\mu_J(\tau)} \Bigr]\, \rho_\syst(0) ,
\end{equation}
where $\mathbf W^\mu_J(\tau)$ are modified influence phase superoperators:
\begin{align}
	&\mathbf W^\mu_J(\tau) \equiv \int_0^\tau \dd\tau'\, \Bigl\{ C_\rett^\mu(\Delta\tau) \bigl[ \tilde Q^\mu_\phantk(\tau)^\times {+} J^\mu_\phantk(\tau) \bigr] \tilde Q^\mu_\phantk(\tau')^\larrow \nonumber\\
	&\qquad - C_\advt^\mu(\Delta\tau) \bigl[ \tilde Q^\mu_\phantk(\tau)^\times {+} J^\mu_\phantk(\tau) \bigr] \bigl[ \tilde Q^\mu_\phantk(\tau')^\rarrow {+} J^\mu_\phantk(\tau') \bigr] \Bigr\}
\end{align}
with $\Delta\tau \equiv \tau - \tau'$.
This result shows that $\tr_\envt \rho_J(t)$ is invariant under replacing equivalent environments, which concludes our proof.

In order to compute the expectation value $E$, one therefore simply has to evaluate Eq.~\eqref{eq:expectation_values} in a pseudomode model.
It is remarkable that both such expectation values and multi-time correlation functions can be computed using pseudomodes in such a straightforward manner:
	one obtains the correct results by basically pretending that the pseudomodes are the real, physical environment.
These results justify our earlier statement, that pseudomodes can often be treated as if they were physical degrees of freedom.
However we note that there are also situations where it is not possible to use pseudomodes as a stand-in for the unitary environment; for example, when calculating local bath occupation numbers.

\subsection{Thermodynamics} \label{subsec:td}

The setup discussed in this paper, consisting of a discrete quantum system and multiple bosonic environments, is commonly used to study quantum thermal machines \cite{AlickiJPhysA1979, QuanPhysRevLett2006, QuanPhysRevE2007, Binder2018, CangemiArXiv230200726Quant-Ph2023, Strasberg2022}.
The open system plays the role of the machine's working medium, and the environments the role of the heat reservoirs.
We will now explore how thermodynamic quantities such as work and heat can be expressed in the pseudomode picture.
For simplicity, we will assume that the system-environment coupling is not altered externally, i.e., that the system coupling operators $Q^\mu$ do not depend on time.

In the strong coupling scenario, the question how to theoretically identify work, heat and related quantities has not yet been universally answered \cite{Strasberg2022}.
Here, we focus on some common, straightforward definitions following Ref.~\cite{KatoThermodynamicsintheQuantumRegime2018}, and use them to demonstrate the application of the pseudomode equivalence.
As a consequence, the following definitions are not meant to be thermodynamically consistent. 
They ignore changes of the interaction energies and can thus violate the first law of thermodynamics.

The internal energy of the working medium can be identified with the expectation value of the system Hamiltonian,
\begin{equation}
	U(t) \equiv \expval{ H_\syst(t) }_\syst .
\end{equation}
Here, we used $\expval{\bullet}_\syst \equiv \tr[ \bullet\, \rho_\syst(t) ]$ for the expectation value with respect to $\rho_\syst(t)$.
Similarly, we will use $\expval{\bullet}$ and $\expval{\bullet}_\CLt$ to denote expectation values in the states $\rho(t)$ and $\rho_\CLt(t)$.
Since the internal energy depends only on the reduced system state, it can obviously be equivalently calculated in the pseudomode model.
The same is true for the power, i.e., the external work performed on the machine per time, which equals the change of the energy due to the external driving:
\begin{equation}
	\dot w(t) \equiv \expval{ \dot H_\syst(t) }_\syst .
\end{equation}

The heat current from the system into a reservoir may be identified with the change of that reservoir's energy:
\begin{equation}
	\dot q_\phantq^\mu(t) \equiv \partial_t \expval{ H_\envt^\mu }_\CLt .
\end{equation}
We note that this definition, sometimes called the \emph{bath heat current} \cite{KatoJChemPhys2016}, guarantees an integrated version of the second law of thermodynamics \cite{EspositoNewJPhys2010, KatoJChemPhys2016} for factorized initial states,
\begin{equation}
	S_\vNt(t) - S_\vNt(0) + \int_0^t \dd\tau \sum\nolimits_\mu \beta^\mu \dot q^\mu(\tau) \geq 0 .
\end{equation}
Here, $S_\vNt(t) \equiv -\expval{ \log \rho_\syst(t) }_\syst$ denotes the von Neumann entropy of the system, with Boltzmann's constant set to one.

In order to translate the heat currents into the pseudomode picture, we must first express them in terms of expectation values of the form \eqref{eq:expectation_values_cl}.
Inserting the unitary time evolution equation of $\rho_\CLt(t)$, one derives \cite{KatoJChemPhys2016}
\begin{equation}
	\dot q^\mu_\phantk(t) = \dot q^\mu_{\systt\phantk}(t) - \partial_t \expval{ H_\intt^\mu }_\CLt + \ii \sum\nolimits_\nu \expval{ \comm{H_\intt^\nu}{H_\intt^\mu} }_\CLt
\end{equation}
where $\dot q^\mu_\systt(t) \equiv \ii \expval{ \comm{H_\syst(t)}{H_\intt^\mu} }_\CLt$ are called system heat currents and the last term may be interpreted as cross currents between the heat reservoirs.
We can now perform the substitution $H^\mu_\intt \to \sum_n H^\mu_{\intt,n}$, with
\begin{equation}
	H^\mu_{\intt,n} \equiv \lambda^\mu_n Q^\mu X^\mu_n
\end{equation}
to obtain an equivalent expression in the pseudomode picture.
This procedure yields $\dot q^\mu(t) = \sum_n \dot q^\mu_n(t)$ with
\begin{align}
	\dot q^\mu_n(t) &\equiv \ii \expval{ \comm{H_\syst^\phantq(t)}{H^\mu_{\intt,n}} } - \partial_t \expval{ H^\mu_{\intt,n} } + \ii \sum\nolimits_{\nu m} \expval{ \comm{H^\nu_{\intt,m}}{H^\mu_{\intt,n}} } \nonumber\\
		&= -\tr\bigl[ H^\mu_{\intt,n}\, \mathbf L^\mu_n\, \rho(t) \bigr] . \label{eq:heat_current_pms}
\end{align}
In the second line, we used the time evolution equation \eqref{eq:pm_evo} of the system-pseudomode state.

The heat currents naturally decompose into a sum of pseudo-currents $\dot q^\mu_n(t)$ associated with the individual pseudomodes.
Their formal expression resembles the heat currents in the standard quantum thermodynamic framework for the Markovian weak-coupling limit, see e.g.\ Refs.~\cite{VinjanampathyContempPhys2016, Binder2018, Kurizki2022, CangemiArXiv230200726Quant-Ph2023}.
We can thus, again, formally treat a pseudomode as if it was physical.
However, it is important to note that the pseudo-currents are not guaranteed to be real-valued.
Since the sum of the pseudo-currents is the physical quantity $\dot q^\mu(t)$, their imaginary parts must cancel each other out.
We will show an example of this phenomenon in Sec.~\ref{sec:examples}.

As shown for example in Refs.~\cite{BrenesPhysRevX2020, MascherpaPhysRevA2020, MedinaPhysRevLett2021}, it is always possible to find a replacement environment consisting of only physical pseudomodes that follow a completely positive time evolution.
Making the pseudomodes physical comes at the cost of more complicated calculations and typically requiring more pseudomodes to match the bath auto-correlation functions with the same accuracy \cite{MascherpaPhysRevA2020, MedinaPhysRevLett2021}.
Such pseudomode models are therefore less suitable for practical calculations, but their mere existence can allow us to infer qualitative properties of the Caldeira-Leggett model.
For example, consider a system without time-dependent driving, that is, with a constant Hamiltonian $H_\CLt$.
Looking only through the lens of unitary system-bath evolution, the long-time behavior of the system and the heat currents would not be immediately apparent.
However, since the system-pseudomode state $\rho(t)$ follows a (regular) Lindblad equation, it will approach a non-equilibrium steady state at long times (given that some basic conditions are satisfied, see e.g.~Refs.~\cite{SpohnLettMathPhys1977, MenczelJPhysA2019}).
We can therefore immediately deduce that the system enters a non-equilibrium steady state at long times, where all interaction observables, such as heat currents, approach steady-state values.

\subsection{Quantum Jump Trajectories} \label{subsec:qjt}

\subsubsection{Background}

The time evolution of the system-pseudomode state formally resembles a Lindblad equation, but it is not completely positive and does not even preserve the Hermiticity of the state.
In the following, we shall consider this type of time evolution equation on a general Hilbert space $\mathcal H$ and temporarily forget about the system-pseudomode structure of our setup.
The evolution has the form
\begin{align}
	\partial_t \rho(t) = &-\ii \comm{H(t)}{\rho(t)} \nonumber\\
		&+ \sum\nolimits_\alpha \gamma^\phantq_\alpha \bigl( L^\phantq_\alpha \rho(t) L_\alpha^\dagger - \acomm{L_\alpha^\dagger L^\phantq_\alpha}{\rho(t)} / 2 \bigr) , \label{eq:general_lindblad}
\end{align}
where $H(t)$ is a Hamiltonian (which may be non-Her\-mitian), and the index $\alpha$ enumerates dissipative channels with corresponding Lindblad operators $L_\alpha$ and rates $\gamma_\alpha$ (which may be complex-valued).

Quantum jump trajectories are a valuable tool for the study of regular Lindblad equations, where $H(t)$ is Hermitian and $\gamma_\alpha$ positive.
It is then possible to write $\rho(t)$ as the statistical average of states $\ket{\psi(t)}$ that evolve according to a stochastic differential equation, i.e., $\rho(t) = \mathbb E\{ \ketbra{\psi(t)}{\psi(t)} \}$.
Hereafter, we call a single realization of such a stochastic differential equation a trajectory, and we use $\mathbb E$ to denote the expectation value in the ensemble of trajectories.

In the present case, the state $\rho(t)$ will generally be non-Hermitian and can therefore not be represented as a statistical average of pure states.
However, master equations such as Eq.~\eqref{eq:general_lindblad} can still be unravelled into stochastic trajectories.
A general technique that achieves this goal has been proposed in Ref.~\cite{BreuerPhysRevA1999}, see also Ref.~\cite{Breuer2002}.
This technique, which applies to generic time-local master equations, considers trajectories on the double Hilbert space $\mathcal H \otimes \mathbb C^2$.
More recently, it was also shown that the dynamics of any time-local master equation can be mapped to an equivalent Lindblad equation on the double Hilbert space \cite{HushPhysRevA2015}.
Motivated by these results, we will now introduce a jump trajectory framework which is based on trajectories in the double Hilbert space and applies specifically to the master equation \eqref{eq:general_lindblad}.

\subsubsection{Unraveling}

The solution of Eq.~\eqref{eq:general_lindblad} can be represented as the statistical average $\rho = \mathbb E\{ \rho_\Psi \}$, where $\Psi \equiv (\ket{\psi_1}, \ket{\psi_2})$ is a trajectory in the double Hilbert space and we set $\rho_\Psi \equiv \ketbra{\psi_1}{\psi_2}$.
Note that we omit all time dependences for brevity.
We require the state $\Psi$ to evolve piecewise deterministically, according to the following Itô stochastic differential equation:
\begin{align}
	\dd \ket{\psi_1} &= \Bigl[ -\ii H - \frac 1 2 \sum\nolimits_\alpha \bigl( \gamma^\phantq_\alpha L_\alpha^\dagger L^\phantq_\alpha - r^\phantq_\alpha \bigr) \Bigr] \ket{\psi_1}\, \dd t \nonumber\\
		&\quad + \sum\nolimits_\alpha \Bigl[ \sqrt{\frac{\gamma_\alpha}{r_\alpha}} L_\alpha \ket{\psi_1} - \ket{\psi_1} \Bigr]\, \dd N_\alpha , \nonumber\displaybreak[0]\\
	\dd \ket{\psi_2} &= \Bigl[ -\ii H^\dagger - \frac 1 2 \sum\nolimits_\alpha \bigl( \gamma_\alpha^{\ast\phantq} L_\alpha^\dagger L_\alpha^\phantq - r_\alpha^\phantq \bigr) \Bigr] \ket{\psi_2}\, \dd t \nonumber\\
		&\quad + \sum\nolimits_\alpha \Bigl[ \sqrt{\frac{\gamma_\alpha}{r_\alpha}}^{\,\ast}\! L_\alpha \ket{\psi_2} - \ket{\psi_2} \Bigr]\, \dd N_\alpha . \label{eq:sde}
\end{align}
Here, the random variables $N_\alpha$ counting the number of jumps in the respective dissipative channels are independent Poisson processes.
Their differentials satisfy the rules $\dd N_\alpha\, \dd N_\beta = \delta_{\alpha\beta}\, \dd N_\alpha$ and $\dd t\, \dd N_\alpha = 0$.
The rates $r_\alpha > 0$, which we keep unspecified for now and may depend on the time and on the state $\Psi$, govern the frequency of jumps:
\begin{equation}
	\mathbb E\{ \dd N_\alpha \mid \Psi \} = r_\alpha\, \dd t .
\end{equation}
This notation stands for the expectation value conditioned on the state $\Psi$ at the beginning of the time step.
Note that for a completely positive (CP) master equation with $H = H^\dagger$ and $\gamma_\alpha \geq 0$, the evolution of $\ket{\psi_1}$ and $\ket{\psi_2}$ is the same.
In this case, we obtain the standard unraveling of the Lindblad equation with the choice
\begin{equation}
	r_\alpha^{\text{CP}} = \gamma_\alpha^\phantq \bra{\psi} L_\alpha^\dagger L_\alpha^\phantq \ket{\psi} ,
\end{equation}
where $\ket{\psi} \equiv \ket{\psi_1} = \ket{\psi_2}$.

To show that the prescription \eqref{eq:sde} reproduces the desired master equation on average, we first apply Itô's lemma,
\begin{equation}
	\dd \rho_\Psi = \left( \dd \ket{\psi_1} \right) \bra{\psi_2} + \ket{\psi_1} \left( \dd \bra{\psi_2} \right) + \left( \dd \ket{\psi_1} \right)\left( \dd \bra{\psi_2} \right) .
\end{equation}
We then plug in the stochastic differential equation and obtain
\begin{align}
	\dd \rho_\Psi &= \Bigl[ -\ii \comm{H}{\rho_\Psi} - \sum\nolimits_\alpha \bigl( \gamma_\alpha^\phantq \acomm{L_\alpha^\dagger L_\alpha^\phantq}{\rho_\Psi} / 2 - r_\alpha^\phantq \rho_\Psi \bigr) \Bigr]\, \dd t \nonumber \\
		&\quad + \sum\nolimits_\alpha \Bigl[ \frac{\gamma_\alpha}{r_\alpha} L_\alpha^\phantq \rho_\Psi L_\alpha^\dagger - \rho_\Psi \Bigr]\, \dd N_\alpha .
\end{align}
Using $\mathbb E\{ f(\Psi)\, \dd N_\alpha \} = \mathbb E\{ f(\Psi)\, r_\alpha(\Psi) \}\, \dd t$, where $f(\Psi)$ is any function of $\Psi$, we can take the expectation value on both sides of this equation.
After some simplifications, one finds that $\rho \equiv \mathbb E\{ \rho_\Psi \}$ satisfies the master equation \eqref{eq:general_lindblad}.

In order to use this unraveling in practice, one must choose a representation of the initial state $\rho(0)$ as a linear combination of states $\rho_\Psi$.
Due to linearity, one can then generate trajectories for each initial state separately and eventually form the linear combination of the results.
In our application to pseudomode models, the initial state is always a normal operator and can therefore be written as $\rho(0) = \sum_i c_i\, \ketbra{ \psi^i }{ \psi^i }$ with $\braket{\psi^i}{\psi^j} = \delta^{ij}$ and $\sum_i c_i = 1$.
We therefore generate trajectories starting from $\ket{\psi_1(0)} = \ket{\psi_2(0)} = \ket{\psi^i}$ for each eigen\-vector $\ket{\psi^i}$ of $\rho(0)$ with corresponding non-zero eigenvalue.
In our examples in Sec.~\ref{sec:examples}, we chose to make the number of trajectories that start from $\ket{\psi^i}$ proportional to $\abs{c_i} / \bigl( \sum_i \abs{c_i} \bigr)$.

Another consideration in practice is how solutions of Eq.~\eqref{eq:sde} can be efficiently numerically sampled.
In our examples, we employed a form of Gillespie's algorithm \cite{GillespieAnnuRevPhysChem2007}, which replaces the sampling of the random variables $\dd N_\alpha$ at each time step $\dd t$ with the inversion sampling of waiting time distributions.
For the reader's convenience, the algorithm is summarized in Appendix \ref{app:gillespie}.

We briefly discuss the relation between our unraveling and earlier publications.
The unraveling proposed in Ref.~\cite{BreuerPhysRevA1999}, applied to the master equation \eqref{eq:general_lindblad}, is a special case of our result with
\begin{equation} \label{eq:mc:breuer}
	r_\alpha^{\text{BKP}} = \abs{\gamma_\alpha}\, \frac{\bra{\psi_1} L_\alpha^\dagger L_\alpha^\phantq \ket{\psi_1} + \bra{\psi_2} L_\alpha^\dagger L_\alpha^\phantq \ket{\psi_2}}{\braket{\psi_1}{\psi_1} + \braket{\psi_2}{\psi_2}} .
\end{equation}
We will here use a different choice of rates, see Eq.~\eqref{eq:mc:rates} below, and compare our choice with Eq.~\eqref{eq:mc:breuer} in Appendix \ref{app:mc:comparison}, showing that our choice gives better convergence.
(The index ``BKP'' stands for the authors of Ref.~\cite{BreuerPhysRevA1999}.)

An important class of non-Markovian dynamics is characterized by master equations of the form \eqref{eq:general_lindblad}, where $H$ is Hermitian and the rates are real-valued but may be negative. This type of dynamics arises, for example, in the study of Redfield equations \cite{DavidovicQuantum2020} or of classical noise \cite{GneitingPhysRevB2020, GroszkowskiQuantum2023}.
In Refs.~\cite{DonvilNatCommun2022, DonvilNewJPhys2023, DonvilOpenSystInfDyn2023}, the authors study master equations of this form and develop a stochastic unraveling that can also be recovered as a special case of ours; we discuss details in Appendix \ref{app:martingale}.
For the same type of master equation, another scheme exists in which temporarily negative rates ``undo'' the effects of earlier jumps \cite{PiiloPhysRevLett2008, MazzolaPhysRevA2009}.
Whether this scheme can be extended to complex-valued rates may be an interesting question for future research.
Other approaches that we will not explore here include the rate operator unraveling for P-divisible dynamics \cite{SmirnePhysRevLett2020} and the use of physical pseudomodes in order to apply the quantum jump unraveling of regular Lindblad equations \cite{ImamogluPhysRevA1994, MazzolaPhysRevA2009}.

\subsubsection{Stability}

In the standard unraveling of regular Lindblad equations, the norm of the state is strictly conserved along each trajectory.
This property is desirable since it contributes to the stability of jump trajectory based simulations of Lindblad equations, for the following reasons.
First, the uncertainty in the result of any computed averages is proportional to the typical size of the contributing ensemble members.
It is therefore beneficial to keep the trajectories bounded.
Furthermore, the norm conservation prevents that some trajectories become negligible over the course of the simulation and thereby reduce the effective size of the ensemble.

This property does not translate to our more general scenario, where $H$ is non-Hermitian or some rates $\gamma_\alpha$ non-positive.
Since the dynamics is trace-preserving, one might expect that the trace $\tr \rho_\Psi = \braket{\psi_2}{\psi_1}$ could play a similar role to that of the norm in a regular Lindblad equation.
However, this quantity cannot be conserved on the trajectory-level, since no choice of $r_\alpha > 0$ makes
\begin{equation}
	\dd( \tr \rho_\Psi ) = \sum\nolimits_\alpha \Bigl( \frac{\gamma_\alpha}{r_\alpha} \expval{ L_\alpha^\dagger L_\alpha^\phantq }_\Psi - \tr \rho_\Psi \Bigr) (\dd N_\alpha - r_\alpha\, \dd t) \label{eq:trace_growth}
\end{equation}
vanish in general.
Here, we used $\expval{\bullet}_\Psi$ to mean $\bra{\psi_2} \bullet \ket{\psi_1}$.
The states along the individual trajectories therefore cannot be normalized to have unit trace; the normalization of the state operator is only recovered in the average.

The variance of the trace in the ensemble of trajectories is given by $\mathbb E\{ \abs{\tr \rho_\Psi}^2 \} - 1$.
To study its behavior, we consider
\begin{align}
	\mathbb E\{ \dd( \abs{\tr \rho_\Psi}^2 ) \mid \Psi \} = \sum\nolimits_\alpha \frac{\abs{\tr \rho_\Psi}^2}{r_\alpha}\, \abs*{r_\alpha - \frac{\gamma_\alpha^\phantq \expval{ L_\alpha^\dagger L_\alpha^\phantq }_\Psi}{\tr \rho_\Psi}}^2 \dd t . \label{eq:mc:fluctuations}
\end{align}
Since no choice of $r_\alpha > 0$ makes this expression vanish, the variance is exponentially increasing in time.
It is thus unavoidable that the absolute value of the trace becomes large on some trajectories.
Note that due to the Cauchy-Schwarz inequality,
\begin{equation} \label{eq:cs}
	\abs{ \tr \rho_\Psi }^2 \leq \braket{\psi_1}{\psi_1} \braket{\psi_2}{\psi_2} ,
\end{equation}
the norm of at least one of the states $\ket{\psi_1}$ and $\ket{\psi_2}$ must then also be large.

These results indicate that the uncertainty in any ensemble average will typically grow exponentially in time.
The unraveling \eqref{eq:sde} is therefore more suitable for short- or intermediate-time simulations than for the study of long-time behavior.
In order to minimize the impact of this effect, we choose $r_\alpha$ such that it minimizes Eq.~\eqref{eq:mc:fluctuations}.
The optimal value is
\begin{equation} \label{eq:mc:rates}
	r_\alpha^\ast = \abs*{ \frac{\gamma_\alpha^\phantq \expval{ L_\alpha^\dagger L_\alpha^\phantq }_\Psi}{\tr \rho_\Psi} } .
\end{equation}
We discuss other choices of rates which, for example, minimize the expected change of the right hand side of Eq.~\eqref{eq:cs} in Appendix \ref{app:qjt}.
There, we also introduce a more general unraveling ansatz, and find that the instability persists.
It thus appears to be a generic feature of unravelings of not-completely positive master equations (and not due to Eq.~\eqref{eq:sde} being badly constructed).

We conjecture that this instability is generic and unavoidable.
Note that the same phenomenon occurs in Refs.~\cite{BreuerPhysRevA1999, DonvilNatCommun2022} since their unravelings are special cases of ours.
A related issue is observed in Ref.~\cite{HushPhysRevA2015}, where the mapping between a non-Lindblad master equation and a Lindblad equation on the double space involves an exponentially growing conversion factor.
Moreover, other stochastic methods lead to similar issues when applied to non-Markovian systems, see for example Refs.~\cite{StockburgerPhysRevLett2002, TanimuraJPhysSocJpn2006, StockburgerEPL2016, TilloyQuantum2017, TanimuraJChemPhys2020, LuoPRXQuantum2023}.

To further support this conjecture, let us consider the single-qubit master equation
\begin{equation}
	\partial_t \rho(t) = \gamma \bigl( \sigma_- \rho(t) \sigma_+ - \acomm{\sigma_+ \sigma_-}{\rho(t)} / 2 \bigr) ,
\end{equation}
where $\sigma_\pm$ are the ladder operators of the qubit and $\gamma \in \mathbb C$.
For $\Re(\gamma) < 0$, the population of the state $\ket +$ is increasing, but the number of trajectories inhabiting that state can only decrease over time.
Therefore, the trajectories that remain in this state must be weighted stronger and stronger to recover the correct solution.
Furthermore, for $\rho(0) = \ketbra + +$, the equation is solved by
\begin{equation}
	\rho(t) = \ee^{-\gamma t}\, \ketbra + + + (1 - \ee^{-\gamma t})\, \ketbra - - .
\end{equation}
We see that for any $\gamma \notin \mathbb R$, the absolute value of the population of the state $\ket -$ will exceed one at some times.
Hence, the state can never be written as a statistical average of normalized states.

\section{Examples} \label{sec:examples}

\subsection{Example 1} \label{subsec:ex1}

\begin{table}
	\begin{enumerate}[(a),itemsep=1em,leftmargin=\parindent,align=left,labelwidth=\parindent,labelsep=0pt]
	\item \adjustbox{valign=t}{
		\begin{tabular*}{0.93\linewidth}{l@{\extracolsep{1cm}}l@{\extracolsep{\fill}}lll}
			\multicolumn{5}{l}{\textbf{Underdamped Environment}} \\
			\toprule
			$n$ & $\Omega_n$ & $\Gamma_n$ & $N_n$ & $\lambda_n^2$ \\
			\midrule
			$1$ & $\Im(\nu_+)$ & $2\gamma$ & $\frac{a_-}{a_+ - a_-^\ast}$ & $a_+ - a_-^\ast$ \\
			$2$ & $0$ & $2\nu_+$ & $0\vphantom{\frac{a_-}{a_+ - a_-^\ast}}$ & $a_+ - a_+^\ast$ \\
			$2 + k$ & $0$ & $2\nu_k$ & $0$ & $a_k$ \\
			\bottomrule
		\end{tabular*}}
	\item \adjustbox{valign=t}{
		\begin{tabular*}{0.93\linewidth}{l@{\extracolsep{1cm}}l@{\extracolsep{\fill}}lll}
			\multicolumn{5}{l}{\textbf{Drude-Lorentz Environment}} \\
			\toprule
			$n$ & $\Omega_n$ & $\Gamma_n$ & $N_n$ & $\lambda_n^2$ \\
			\midrule
			$1$ & $\frac{1}{2\ii} (\gamma - \Omega)$ & $\gamma + \Omega$ & $0\vphantom{\frac{a_-}{a_+ {-} a_-^\ast}}$ & $a_0$ \\
			$2$ & $\frac{1}{2\ii} (\Omega - \gamma)$ & $\gamma + \Omega$ & $0\vphantom{\frac{a_-}{a_+ {-} a_-^\ast}}$ & $a_0^\ast$ \\
			$2 + k$ & $0$ & $2\nu_k$ & $0$ & $a_k$ \\
			\bottomrule
		\end{tabular*}}
	\end{enumerate}
	\caption[x]{
		Parameters for pseudomodes that are equivalent to an \begin{enumerate*}[(a)]\item underdamped environment [Eq.~\eqref{eq:underdamped_sd}], \item overdamped (Drude-Lorentz) environment [Eq.~\eqref{eq:overdamped_sd}].\end{enumerate*}
		The third line of each table refers to the Matsubara modes with $1 \leq k \leq k_{\text{max}}$, with the corresponding pseudomodes indexed by $n = 2+k$.
		The parameters appearing in the tables are defined in Eqs.~\eqref{eq:underdamped_cfct} to \eqref{eq:underdamped_cfct3} and Eqs.~\eqref{eq:overdamped_cfct} to \eqref{eq:overdamped_cfct3}, respectively.
		The regularization constant $\Omega$ should be chosen larger than all other relevant parameters.
	}
	\label{tab:examples}
\end{table}

In this example, we consider the thermalization of a qubit in an underdamped environment.
We assume that the system Hamiltonian is $H_\syst = (\Delta / 2)\, \sigma_x$, where $\Delta$ is the level splitting and $\sigma_x$ the Pauli matrix, and that it couples to the environment via $Q = \sigma_z$.
At the initial time $t = 0$, the qubit is prepared in the maximally mixed state and brought into contact with a heat reservoir.
The reservoir is characterized by the spectral density of underdamped Brownian motion,
\begin{equation} \label{eq:underdamped_sd}
	G(\omega) = \frac{2\lambda^2 \gamma \omega}{(\omega^2 - \omega_0^2)^2 + 4\gamma^2\omega^2} \Theta(\omega) ,
\end{equation}
and its inverse temperature $\beta$, which we assume to be finite.
Here, $\lambda$ is the coupling strength, $\omega_0$ the reservoir characteristic frequency, $\gamma$ the half-width of the spectral density (with $\gamma < \omega_0$), and $\Theta$ the Heaviside function.

The corresponding reservoir correlation function can be determined from Eq.~\eqref{eq:cl_cfct}; it is
\begin{equation} \label{eq:underdamped_cfct}
	C(t) = a_+\, \ee^{-\nu_+ t} + a_-\, \ee^{-\nu_- t} + \sum_{k=1}^\infty a_k\, \ee^{-\nu_k t} .
\end{equation}
It consists of a resonant contribution with frequencies $\nu_\pm \equiv \gamma \pm \ii (\omega_0^2 - \gamma^2)^{1/2}$ and corresponding coefficients
\begin{equation} \label{eq:underdamped_cfct2}
	a_\pm \equiv \frac{\lambda^2}{4\Im(\nu_\pm)} \Bigl[ 1 + \ii\, \cot \Bigl( \frac{\beta\nu_\pm}{2} \Bigr) \Bigr] ,
\end{equation}
and of a sum of Matsubara terms.
The Matsubara frequencies and coefficients are $\nu_k \equiv 2\pi k / \beta$ and
\begin{equation} \label{eq:underdamped_cfct3}
	a_k \equiv -\frac{4\lambda^2\gamma}{\beta} \frac{\nu_k}{(\nu_+^2 + \nu_k^2)(\nu_-^2 + \nu_k^2)} ,
\end{equation}
respectively.
In practical calculations, one can only include a finite number of Matsubara terms with $k \leq k_{\text{max}}$.
It would be possible to approximate the remaining terms as a $\delta$-contribution to the correlation function which could then be included as a Tanimura terminator like discussed in Appendix \ref{app:influence_generalizations}.
For simplicity, we will however not include such a terminator term here and ignore all terms with $k > k_{\text{max}}$.

We are now ready to construct an equivalent pseudomode environment.
To apply Table~\ref{tab}, we note that $a_k \in \mathbb R$ for the Matsubara terms, and that $\nu_- = \nu_+^\ast$, $(a_+ + a_-) \in \mathbb R$, and $\abs{\Re(a_+)} > \abs{\Re(a_-)}$.
The underdamped environment can thus be represented by the $(k_{\text{max}} + 2)$ pseudomodes listed in Tab.~\ref{tab:examples}(a).
This representation is exact except for the ignored Matsubara modes.
The only other approximation that we make is truncating the pseudomode Hilbert spaces, discarding states with more excitations than a cutoff $C_n$ (where $n$ indexes the pseudomodes).
We note that earlier attempts to model finite-temperature underdamped environments with pseudomodes required at least $(k_{\text{max}} + 3)$ pseudomodes, see for example Appendix D2 in Ref.~\cite{LuoPRXQuantum2023}.

\begin{figure}
	\centering
	\includegraphics[scale=1]{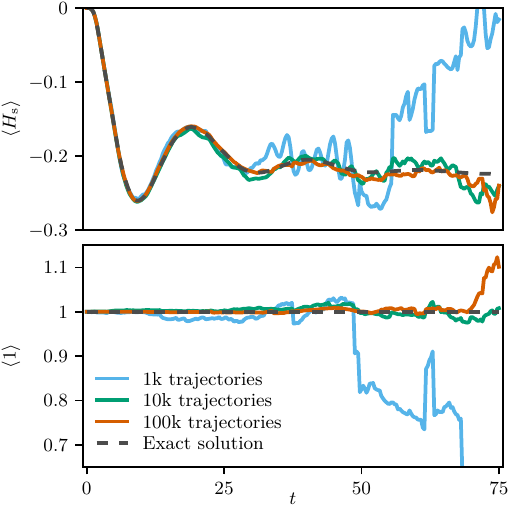}
	\caption{
		Time evolution of a qubit thermalizing in an underdamped environment (Example 1).
		The upper panel shows the behavior of $\expval{H_\syst}$, with the dashed gray curve corresponding to the exact solution obtained by either integrating the pseudo-Lindblad equation or the HEOM.
		The solid curves are Monte Carlo estimates of $\expval{H_\syst}$ generated from a varying number of trajectories following the unraveling \eqref{eq:sde}.
		The lower panel shows Monte Carlo estimates of $\expval 1$.
		That is, the solid curves are obtained by taking the average of $\tr \rho_\Psi$ over the trajectories $\Psi(t)$.
		The deviations of these estimates from $1$ are indicators for the convergence of the Monte Carlo simulations.
		All Monte Carlo curves in this figure show only the real parts.
		We used the parameters $\Delta = 1$, $\lambda = 0.2$, $\gamma = 0.025$, $\omega_0 = 1$ and $\beta = 1$, and the jump rates $r_\alpha^\ast$ defined in Eq.~\eqref{eq:mc:rates}.
	}
	\label{fig:ex1:trajectories}
\end{figure}

The system parameters used in our simulations can be found underneath Fig.~\ref{fig:ex1:trajectories}.
By varying the number of Matsubara exponents, we verified that it is sufficient to only consider the resonant contribution, i.e., to set $k_{\text{max}} = 0$.
We further verified that a truncation with cutoff $C_1 = 9$ and $C_2 = 3$ perfectly reproduces results obtained with a HEOM simulation.
Our code was written in Python using QuTiP \cite{JohanssonComputPhysCommun2012, JohanssonComputPhysCommun2013, LambertPhysRevRes2023} and is available on GitHub, see Ref.~\footnote{\url{https://github.com/pmenczel/Pseudomode-Examples}\label{fn:github}}.

The upper panel of Fig.~\ref{fig:ex1:trajectories} shows the time dependence of the expectation value of the system Hamiltonian together with the results of a Monte Carlo simulation.
We find very good agreement between the exact result and the Monte Carlo simulation at short times, and strong fluctuations in the Monte Carlo results at long times.
As expected from our stability discussions, these fluctuations are hard to tame even with a greatly increased number of trajectories.
Further analysis in Appendix \ref{app:stability} confirms that the time interval on which the Monte Carlo simulation converges grows logarithmically with the number of trajectories.

Since $\tr \rho_\Psi$ is only constant on average and not along trajectories, it is interesting to also consider the Monte Carlo estimate of $\expval 1$, i.e., the ensemble average of $\tr \rho_\Psi$, shown in the bottom panel of Fig.~\ref{fig:ex1:trajectories}.
We find qualitatively the same convergence behavior as for the estimate of $\expval{H_\syst}$.
If the exact solution was unknown, the deviation of the estimate of $\expval 1$ from $1$ could thus be used to judge the convergence of the Monte Carlo simulation.

\begin{figure}
	\centering
	\includegraphics[scale=1]{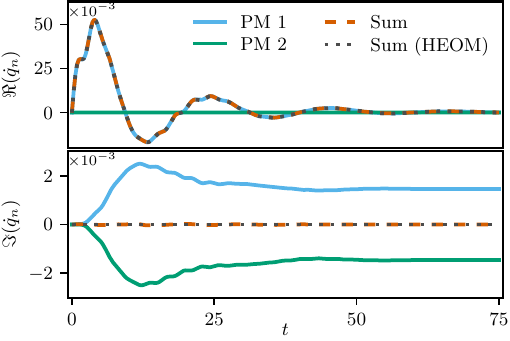}
	\caption{
		Heat currents in Example 1.
		Following Eq.~\eqref{eq:heat_current_pms}, the total heat current $\dot q$ decomposes into contributions $\dot q_n$ associated with the pseudomodes (PMs).
		The solid curves in the upper (lower) panel show the real (imaginary) parts of these contributions, and the dashed orange curves their sums.
		The dotted gray curves are a HEOM calculation based on the method described in Refs.~\cite{KatoJChemPhys2016, LambertPhysRevRes2023}, and they agree well with the sum of the pseudomode heat currents.
		In the upper panel, the curves for the first pseudomode, the sum and the HEOM result all overlap.
	}
	\label{fig:ex1:heat_currents}
\end{figure}

In Fig.~\ref{fig:ex1:heat_currents}, we show the complex-valued heat currents associated with the two pseudomodes according to Eq.~\eqref{eq:heat_current_pms}.
The figure shows that the contribution of the second pseudomode has a vanishing real part; this pseudomode only serves to cancel out the imaginary part of the first pseudomode's contribution.
What remains is the real part of that contribution, which exactly matches the result of a HEOM calculation.
We are thus reminded that intermediate steps of calculations in the pseudomode framework can yield complex-valued values that appear unphysical.
These unphysical values must however combine to the correct physical result in the end.

\subsection{Example 2} \label{subsec:ex2}

In this example, we will study the dynamical decoupling of a qubit from an overdamped Drude-Lorentz environment.
The idea of dynamical decoupling is that the periodic application of $\pi$-pulses to the qubit can counteract the effect of the environment on the qubit \cite{ViolaPhysRevA1998}.
The following setup is identical to an example that was studied in Ref.~\cite{LambertPhysRevRes2023} using the HEOM; we will here use it to demonstrate the applicability of the pseudomode method to overdamped environments and time-dependent driving and add an analysis of the qubit-environment correlations.% and of two-time correlation functions of the qubit.

We work in an interaction picture where the qubit Hamiltonian is
\begin{equation}
	H_\syst(t) = f(t)\, \sigma_x ,
\end{equation}
with $f(t)$ describing the periodic $\pi$-pulses.
Specifically, $f(t) \equiv V$ whenever $t \in [n\tau - \tau_p, n\tau]$ for some integer $n$, and $f(t) \equiv 0$ otherwise.
Here, $V$ is the pulse strength, $\tau^{-1}$ the pulse frequency, and $\tau_p \equiv \pi / (2V)$ the pulse duration.
The qubit is initially in the $\ket{+}_x$ eigenstate of $\sigma_x$ with $\sigma_x \ket{+}_x = \ket{+}_x$.
It couples to the environment with the coupling operator $Q = \sigma_z$.
We assume the spectral density of the environment to be
\begin{equation} \label{eq:overdamped_sd}
	G(\omega) = \frac{2\lambda\gamma\omega}{\gamma^2 + \omega^2} \Theta(\omega) ,
\end{equation}
where $\lambda$ is the coupling strength and $\gamma$ the cutoff frequency \footnote{Following the usual convention where for the overdamped environment, $\lambda$ has units of frequency, whereas for the underdamped environment, $\lambda$ has units of frequency to the power $3/2$.}.

The corresponding environment correlation function is
\begin{equation} \label{eq:overdamped_cfct}
	C(t) = a_0\, \ee^{-\gamma t} + \sum_{k=1}^\infty a_k\, \ee^{-\nu_k t} .
\end{equation}
Again, it consists of a resonant contribution and a sum of Matsubara terms.
The Matsubara frequencies and coefficients are again all real-valued; they are given by $\nu_k \equiv 2\pi k / \beta$ and
\begin{equation} \label{eq:overdamped_cfct2}
	a_k \equiv \frac{4\lambda\gamma\nu_k}{\beta(\nu_k^2 - \gamma^2)} .
\end{equation}
The single remaining coefficient,
\begin{equation} \label{eq:overdamped_cfct3}
	a_0 \equiv \lambda\gamma \Bigl[ \cot\Bigl( \frac{\beta\gamma}{2} \Bigr) - \ii \Bigr] ,
\end{equation}
has a non-zero imaginary part.
We must therefore add a regularization term $a_0^\ast\, \ee^{-\Omega t}$ with $\Omega \gg 1$ to the correlation function.
We then obtain a representation of the environment with $(k_{\text{max}} + 2)$ pseudomodes [listed in Table~\ref{tab:examples}(b)], where $k_{\text{max}}$ is the number of included Matsubara terms.

\begin{figure}
	\centering
	\includegraphics[scale=1]{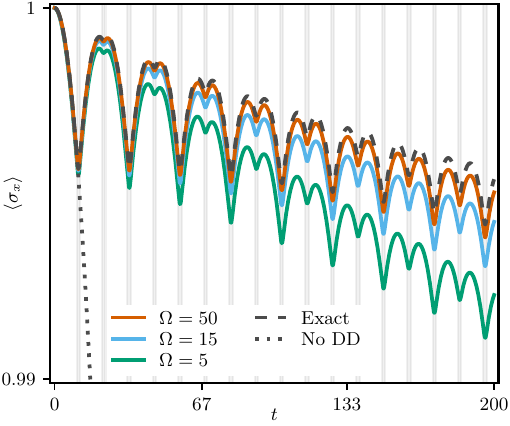}
	\caption{
		Dynamical Decoupling (Example 2).
		The plot shows that with dynamical decoupling, the expectation value $\expval{\sigma_x}$ remains approximately constant instead of decaying.
		The dashed and dotted gray curves are, respectively, the result of HEOM simulations with dynamical decoupling ($H_\syst(t)$ as described in the main text) and without ($H_\syst(t) = 0$).
		The solid curves are pseudomode calculations using different values of the regularization constant $\Omega$, showing that the pseudomode results approach the HEOM result for large $\Omega$.
		The gray shading of the background indicates the times where $f(t) = V$.
		We chose the parameters like in Ref.~\cite{LambertPhysRevRes2023} ($\beta = 10 / V$, $\gamma = 10^{-2}\, V$, $\lambda = 10^{-4}\, V$, $\tau = \tau_p + 10 / V$), and we set $V = 1$.
		The dashed and dotted gray curves are therefore identical to the solid green and dashed orange curves in Fig.~8 there.
	}
	\label{fig:ex2:time_evo}
\end{figure}

In Fig.~\ref{fig:ex2:time_evo}, we verify that this regularization procedure is working by comparing our results with the HEOM calculation of Ref.~\cite{LambertPhysRevRes2023}.
We included $k_{\text{max}} = 3$ Matsubara terms and used the cutoff $C_n = 3$ for all pseudomodes.
For $\Omega = 50\, V$, our results are in good agreement with the HEOM results, and will use this value in the following.
As for the previous example, our code is available on GitHub~\cite{Note1}.

We now want to study the rise and decay of correlations and entanglement between the qubit and its environment.
In a usual bipartite system, a (Hermitian, positive semi-definite) state $\rho$ is called separable if it can be written as a convex combination of product states \cite{GuhnePhysRep2009},
\begin{equation} \label{eq:separable}
	\rho = \sum\nolimits_i p_i\, \rho_{Ai} \otimes \rho_{Bi} ,
\end{equation}
where $p_i \geq 0$ and $\rho_{Ai}$ and $\rho_{Bi}$ are states of the constituent systems.
Otherwise, the state is entangled, and the amount of entanglement can be characterized by various measures including the negativity \cite{GuhnePhysRep2009}
\begin{equation} \label{eq:negativity}
	\mathcal N(\rho) \equiv \frac{\norm{\rho^{T_A}}_{\text{tr}} - 1}{2} = \frac{\sum_{\lambda \in \sigma(\rho^{T_A})} \abs{\lambda} - 1}{2} ,
\end{equation}
which is zero if (but not only if) $\rho$ is separable.
Here, $\rho^{T_A}$ denotes the partial transpose of $\rho$ with respect to the subsystem $A$, $\norm{\bullet}_{\text{tr}}$ the trace norm, and the sum runs over the eigenvalues of $\rho^{T_A}$.

In our case where $\rho$ may be non-Hermitian, the distinction between separable and entangled states is less clear.
Applying the operator Schmidt decomposition \cite{GuhnePhysRep2009}, any operator $\rho$ can be written in the form \eqref{eq:separable} with positive $p_i$ and some operators $\rho_{Ai}$ and $\rho_{Bi}$; the crucial condition for separability is that $\rho_{Ai}$ and $\rho_{Bi}$ must be states, that is, Hermitian and positive semi-definite.
However, a non-Hermitian state $\rho$ can never be written in this form with positive $p_i$ and Hermitian $\rho_{Ai}$ and $\rho_{Bi}$.

\begin{figure}
	\centering
	\includegraphics[scale=1]{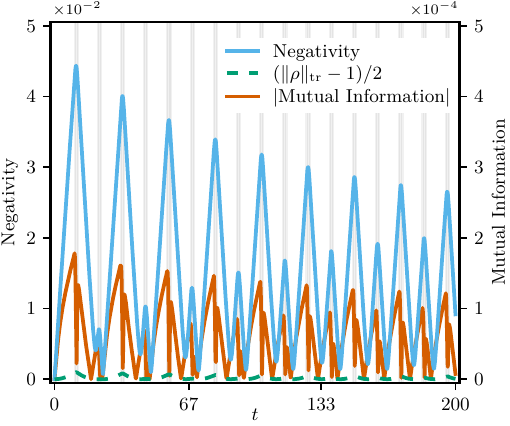}
	\caption{
		System-environment correlations in the dynamical decoupling example.
		The solid blue and red curves respectively show the negativity \eqref{eq:negativity_again} and the absolute value of the quantum mutual information \eqref{eq:qmi} with values on the left and right $y$-axes.
		To compare with the negativity, the dashed green curve shows the expression $(\norm{\rho}_{\text{tr}} - 1) / 2$ which would be zero for a usual (Hermitian, positive semi-definite) state, also with values on the left axis.
		The gray shading of the background indicates the times where $f(t) = V$.
	}
	\label{fig:ex2:entanglement}
\end{figure}

Despite this issue, let us explore the behavior of the negativity of the system-pseudomode state.
For non-Hermitian $\rho$ the second equality of Eq.~\eqref{eq:negativity} does not hold; we will still use
\begin{equation} \label{eq:negativity_again}
	\mathcal N(\rho) \equiv \frac{\norm{\rho^{T_\syst}}_{\text{tr}} - 1}{2}
\end{equation}
in this case (with $\rho^{T_\syst}$ being the partial transpose with respect to the open system).
Figure~\ref{fig:ex2:entanglement} shows that the negativity between the system and the pseudomode environment builds up until the first $\pi$-pulse is applied.
After the $\pi$-pulse, the initial state of the qubit is approximately restored and the negativity returns to zero.
This behavior of the negativity matches the intuition that the qubit-environment correlations form a ``memory'' which can be drawn upon to restore the initial qubit state.
Even though there is no obvious formal relationship between $\mathcal N(\rho)$ here and the actual qubit-environment entanglement, and despite the issues detailed above, our results suggest that $\mathcal N(\rho)$ still provides at least a qualitative indicator of the entanglement.
Therefore we can, again, treat certain properties of the sum of pseudomodes almost as if they pertain to a physical environment.

In addition to the negativity, Fig.~\ref{fig:ex2:entanglement} also shows the quantum mutual information
\begin{equation} \label{eq:qmi}
	I_{\syst : \text{pm}}(\rho) = S(\rho_\syst) + S(\rho_{\text{pm}}) - S(\rho) ,
\end{equation}
where $\rho_\syst$ and $\rho_{\text{pm}}$ denote the partial states of the qubit and the pseudomodes, and $S$ the von Neumann entropy.
For usual states, the quantum mutual information quantifies both the quantum and the classical correlations of the state.
In our case, the quantum mutual information becomes complex-valued and we consider its absolute value, which exhibits a behavior similar to the negativity.
At greater times, the periodicity of the state changes from $2\tau$ to $\tau$; the initially broken time translation symmetry by $\tau$ is thus restored.
Interestingly, this symmetry restoration process seems to happen on a much faster time scale than the overall relaxation, which will eventually bring the system into a $\tau$-periodic limit cycle despite the dynamical decoupling.

\section{Concluding Perspectives} \label{sec:end}

Understanding the interactions of quantum systems with their environment in the non-Markovian and strong-coupling regimes is crucial for the development of quantum technology, for the study of quantum thermodynamics, and for our comprehension of the quantum world in general.
In this work, we have demonstrated that the pseudomode technique may provide a significant contribution to this understanding.
By more formally solidifying the theoretical framework that the technique is built on, we have shown that, despite its apparent unphysicality, a pseudomode environment can be used \emph{in lieu} of the actual one to analyze a multitude of quantities from multi-time correlation functions to quantum trajectories and system-environment currents and correlations.

Furthermore, we have provided in Table~\ref{tab} a handy recipe for the translation of any given environment with a multi-exponential auto-correlation function into a mathematically equivalent pseudomode environment obeying a quantum master equation.
We have demonstrated the application of this recipe on the examples of underdamped Brownian environments and overdamped Drude-Lorentz ones, and we demonstrated that finite-temperature underdamped environments can be described using fewer pseudomodes than what was previously thought possible.

In this work, we have focused on harmonic pseudomodes, that is, such whose free time evolution follows the well-known Lindblad equation for a damped harmonic oscillator -- albeit with complex-valued parameters -- since they are the most straightforward implementation of the general non-unitary environments discussed at the beginning.
The general framework we have introduced also encompasses setups with multiple coupled pseudomodes like in, e.g., Refs.~\cite{MascherpaPhysRevA2020, MedinaPhysRevLett2021, Sanchez-BarquillaNanophotonics2022, LednevPhysRevLett2024}, as well the dissipaton approach \cite{WangJChemPhys2022, LiArXiv240117255Quant-Ph2024}, whose relationship with pseudomodes had not been fully understood until now.
Whether it is possible to find equivalent environments that are not based on harmonic modes, and whether our technique can be generalized even further are both interesting questions for future studies.
Further generalizations might involve non-factorizing initial conditions, treated either like in Eq.~\eqref{eq:thermal_corr_fct} or with the approaches introduced in Refs.~\cite{SmithPhysRevA1987, GrabertPhysRep1988}, or they might even go beyond the complex-valued parameters discussed here and introduce, for example, non-commuting numbers.

By demonstrating how pseudomodes can improve dissipative state engineering algorithms \cite{LambertArXiv231012539Quant-Ph2023}, pseudomodes have already proven to be useful for practical applications.
In Ref.~\cite{BrenesPhysRevB2023}, it was shown that pseudomodes can be used to study fluctuating quantities in strongly-coupled systems.
Our investigations of quantum jump trajectories is a step towards advancing this line of research, and our results thus pave the way for a variety of further applications.
These applications include the study of quantum many-body systems, of the performance of quantum thermal machines, or of fundamental relationships involving the fluctuations of such systems.

\begin{acknowledgments}
We thank B.~Donvil, C.~Flindt, C.~Gneiting, P.~Muratore-Ginanneschi and V.~Vitale for insightful discussions.
We thank G.~Suarez for his careful reading of our manuscript.
We thank the anonymous reviewers for their many very helpful suggestions.

We acknowledge the Information Systems Division, RIKEN, for the use of their facilities.
PM performed this work as an International Research Fellow of the Japan Society for the Promotion of Science (JSPS).
KF acknowledges support from JSPS KAKENHI (Grant Number 23K13036) and MEXT Quantum Leap Flagship Program (Grant Number JPMXS0120330644).
MC acknowledges support from NSFC (Grant No.\ 11935012) and NSAF (Grant No.\ U2330401).
NL acknowledges support from the RIKEN Incentive Research Program.
FN is supported in part by:
	Nippon Telegraph and Telephone Corporation (NTT) Research,
	the Japan Science and Technology Agency (JST) [via the Quantum Leap Flagship Program (Q-LEAP), and the Moonshot R{\&}D Grant Number JPMJMS2061],
	the Asian Office of Aerospace Research and Development (AOARD) (via Grant No.\ FA2386-20-1-4069),
	and the Office of Naval Research (ONR) Global (via Grant No.\ N62909-23-1-2074).
\end{acknowledgments}

\appendix

% to make, for example, \ref{app:influence_functional} into "A1" instead of "A 1"
\makeatletter
\renewcommand{\p@subsection}{\Alph{section}}
\makeatother

\section{Influence Functionals for Non-Unitary Environments} \label{app:equivalence}

\subsection{Derivation} \label{app:influence_functional}

Here, we will derive the influence functional representation \eqref{eq:influence_functional} of the open system dynamics from the time evolution equation \eqref{eq:pm_evo}, which we repeat here:
\begin{equation} \label{eq:app:pm_evo}
	\partial_t \rho(t) = (\mathbf L_\syst(t) + \mathbf L_\envt)\, \rho(t) - \ii \sum\nolimits_{\mu n} \lambda^\mu_n\, \comm[\big]{Q^\mu_\phantq(t) X^\mu_n}{\rho(t)} .
\end{equation}
Here, we set $\mathbf L_\syst(t) \equiv -\ii H_\syst(t)^\times$ and $\mathbf L_\envt \equiv \sum_{\mu n} \mathbf L^\mu_n$.
Recall that for any operator $A$, we define the superoperators $A^\larrow\, \bullet \equiv \bullet\, A$, $A^\rarrow\, \bullet \equiv A\, \bullet$, and $A^\times \equiv A^\rarrow - A^\larrow$.

We move to an interaction picture with respect to the free evolution $\mathbf L_\syst(t) + \mathbf L_\envt$.
For an operator $A(t)$ that may have an explicit time-dependence, we thus define
\begin{align}
	\tilde A(t) &\equiv \timeordinv\Bigl[ \ee^{-\int_0^t\! \dd\tau\, \mathbf L_\syst(\tau)} \Bigr]\, \ee^{-\mathbf L_\envt t} A(t) \nonumber\\
		&= \timeordinv\Bigl[ \ee^{\ii \int_0^t\! \dd\tau\, H_\syst(\tau)} \Bigr] \bigl[ \ee^{-\mathbf L_\envt t} A(t) \bigr]\, \timeord\Bigl[ \ee^{-\ii \int_0^t\! \dd\tau\, H_\syst(\tau)} \Bigr] ,
\end{align}
where $\timeordinv$ denotes inverse time-ordering with later times moved to the right.
The time evolution equation in the interaction picture reads $\partial_t \tilde \rho(t) = \tilde{\mathbf L}_\intt(t) \tilde \rho(t)$, with
\begin{equation}
	\tilde{\mathbf L}_\intt(t) \equiv -\ii \sum\nolimits_{\mu n} \lambda^\mu_n\, \bigl[ \tilde Q_\phantq^\mu(t)^\rarrow (X^\mu_n)^\rarrow_t - \tilde Q_\phantq^\mu(t)^\larrow (X^\mu_n)^\larrow_t \bigr] ,
\end{equation}
using the notation defined in Eq.~\eqref{eq:supop_interaction_pic}.

Let us introduce the following correlation functions:
\begin{align}
	C^\mu_{\advt,n}(t) &\equiv \tr\bigl[ X^\mu_n(t) X^\mu_n\, \rho^\mu_{\eqt,n} \bigr] \quad \text{and} \nonumber\\
	C^\mu_{\rett,n}(t) &\equiv \tr\bigl[ X^\mu_n X^\mu_n(t)\, \rho^\mu_{\eqt,n} \bigr] . \label{eq:app:pm_cfcts}
\end{align}
Here, $X^\mu_n(t)$ is the Heisenberg picture operator as defined in Eq.~\eqref{eq:heisenberg}.
Using our assumptions that $\mathbf L^\mu_n$ is trace-preserving and that $\mathbf L^\mu_n \rho^\mu_{\eqt,n} = 0$, we derive the identities
\begin{align}
	\tr\bigl[ (X^\mu_n)^i_\tau (X^\mu_n)^\rarrow_{\tau'} \rho^\mu_{\eqt,n} \bigr] &= C^\mu_{\advt,n}(\tau - \tau') \quad \text{and} \nonumber\\
	\tr\bigl[ (X^\mu_n)^i_\tau (X^\mu_n)^\larrow_{\tau'} \rho^\mu_{\eqt,n} \bigr] &= C^\mu_{\rett,n}(\tau - \tau') \label{eq:app:corr_derivation}
\end{align}
for any $i \in \{ \larrow, \rarrow \}$.

We are now ready to calculate the influence functional, proceeding along the lines of the unitary case explained in Ref.~\cite{Breuer2002}.
We formally write the system state as
\begin{align}
	\tilde \rho_\syst(t) &= \tr_\envt\biggl\{ \timeord\exp \biggl[ \int_0^t \dd\tau\, \tilde{\mathbf L}_\intt(\tau) \biggr] \rho(0) \biggr\} \nonumber\\
		&= \timeord_\syst \tr_\envt\biggl\{ \timeord_\envt\exp \biggl[ \int_0^t \dd\tau\, \tilde{\mathbf L}_\intt(\tau) \biggr] \rho(0) \biggr\} ,
\end{align}
where $\tr_\envt$ denotes the partial trace over all of the auxiliary environment.
The time-ordering operators $\timeord_\syst$ and $\timeord_\envt$ act only on the system-space expressions $\tilde Q^\mu(t)^i$ and environment-space expressions $(X^\mu_n)^i_t$, respectively.
Since all commutators $\comm{(X^\mu_n)^i_t}{(X^\mu_n)^j_{t'}}$ are complex numbers by our third assumption, $\comm{\tilde{\mathbf L}_\intt(\tau)}{\tilde{\mathbf L}_\intt(\tau')}$ acts only on the system space.
Applying Wick's theorem \cite{Breuer2002}, we thus arrive at
\begin{equation} \label{eq:stuff1}
	\vspace{.5em} % small layouting fix
	\tilde \rho_\syst(t) = \timeord_\syst\Bigl\{ \ee^{\frac 1 2 \int_0^t\! \dd\tau\! \int_0^\tau\! \dd\tau'\, \comm{\tilde{\mathbf L}_\intt(\tau)}{\tilde{\mathbf L}_\intt(\tau')}} \tr_\envt\Bigl[ \ee^{\int_0^t\! \dd\tau\, \tilde{\mathbf L}_\intt(\tau)} \rho(0) \Bigr] \Bigr\} .
	\vspace{.5em} % small layouting fix
\end{equation}

We are assuming that the initial state factorizes as $\rho(0) = \rho_\syst(0) \otimes \rho_\envt(0)$ with Gaussian $\rho_\envt(0)$.
By ``Gaussian'', we mean that all $n$-point correlators vanish for odd $n$ and decompose into $2$-point correlators for even $n$.
We can formally express this assumption as
\begin{align} \label{eq:app:wick_theorem}
	&(2^n n!) \expval[\big]{ \mathbf A_1 \cdots \mathbf A_{2n} } \nonumber\\
	&= \sum\nolimits_\sigma \expval[\big]{ \mathcal N\bigl[ \mathbf A_{\sigma(1)} \mathbf A_{\sigma(2)} \bigr] } \cdots \expval[\big]{ \mathcal N\bigl[ \mathbf A_{\sigma(2n{-}1)} \mathbf A_{\sigma(2n)} \bigr] } ,
\end{align}
where $\expval{\bullet} \equiv \tr[ \bullet\, \rho_\envt(0) ]$, the superoperators $\mathbf A_k$ all have the form $(X^\mu_n)^i_t$, the sum runs over all permutations of the $2n$ indices, and $\mathcal N$ denotes the natural ordering, i.e., it moves superoperators with smaller indices to the left.
Making use of this identity and of the fact that the factors $\tilde Q^\mu(t)^i$ behave like commuting numbers inside the time ordering $\mathcal T_\syst$, we can write the system state as follows:
\begin{widetext}
	\begin{equation} \label{eq:stuff2}
		\tilde \rho_\syst(t) = \timeord_\syst\biggl[ \exp\biggl\{ \frac 1 2 \int_0^t \dd\tau \int_0^\tau \dd\tau'\, \comm{\tilde{\mathbf L}_\intt(\tau)}{\tilde{\mathbf L}_\intt(\tau')} \biggr\}\, \exp\biggl\{ \frac 1 2 \int_0^t \dd\tau \int_0^t \dd\tau'\, \expval{ \tilde{\mathbf L}_\intt(\tau) \tilde{\mathbf L}_\intt(\tau') }_\envt \biggr\}\, \rho_\syst(0) \biggr] .
	\end{equation}
\end{widetext}
We finally restore the Schrödinger picture and obtain
\begin{equation} \label{eq:app:inf_functional}
	\rho_\syst(t) = \timeord\Bigl[ \ee^{\int_0^t \dd\tau\, \mathbf L_\syst(\tau)} \Bigr]\, \timeord\Bigl[ \ee^{\int_0^t \dd\tau\, \mathbf W(\tau)} \Bigr]\, \rho_\syst(0)
\end{equation}
with the superoperator
\begin{equation} \label{eq:app:fv_derivation}
	\mathbf W(\tau) \equiv \int_0^\tau \dd\tau'\, \expval[\big]{ \tilde{\mathbf L}_\intt(\tau) \tilde{\mathbf L}_\intt(\tau') }_\envt .
\end{equation}
Using some straightforward algebra and the identities \eqref{eq:app:corr_derivation}, one can see that $\mathbf W(\tau) = \sum_\mu \mathbf W^\mu(\tau)$ with the influence phase superoperators $\mathbf W^\mu(\tau)$ defined in Eq.~\eqref{eq:influence_phase}.
We have thus derived Eq.~\eqref{eq:influence_functional}.

In the derivation of Eq.~\eqref{eq:mtcf_proof}, a source term is added to the interaction term $\tilde{\mathbf L}_\intt(t)$.
With this source term, equality between Eqs.~\eqref{eq:stuff1} and \eqref{eq:stuff2} does not hold.
However, the source term acts trivially on the environment subspace; it can therefore be pulled out of the partial trace in Eq.~\eqref{eq:stuff1}.
Afterwards, we can proceed as above to arrive at the desired result.
In the derivation of Eq.~\eqref{eq:interaction_proof}, the source terms are linear in the $(X^\mu_n)^i_t$ and, hence, Eq.~\eqref{eq:stuff2} holds without modification here.
The result is obtained by plugging the modified $\tilde{\mathbf L}_\intt(t)$ into this equation.

\subsection{Generalizations} \label{app:influence_generalizations}

So far, we have focused on the relatively simple evolution equation \eqref{eq:pm_evo}.
We briefly discuss some straightforward generalizations.
To keep the presentation simple, we will consider each generalization separately, but they can be easily combined.

\subsubsection*{Multiple Coupling Terms}
First, we consider multiple coupling terms per environment.
The time evolution equation then becomes $\partial_t \rho(t) = (\mathbf L_\syst(t) + \mathbf L_\intt(t) + \mathbf L_\envt) \rho(t)$ with an interaction term
\begin{equation}
	\mathbf L_\intt(t)\, \rho \equiv -\ii \sum\nolimits_{\mu n \alpha} \lambda^\mu_{n\alpha}\, \comm[\big]{Q^\mu_\alpha(t) X^\mu_{n \alpha}}{\rho} .
\end{equation}
The calculation proceeds exactly like above until we reach Eq.~\eqref{eq:app:fv_derivation}.
Plugging in the new interaction term, we find $\mathbf W(\tau) = \sum_{\mu \alpha \beta} \mathbf W^\mu_{\alpha\beta}(\tau)$ with
\begin{align}
	\mathbf W^\mu_{\alpha\beta}(\tau) \equiv &- \int_0^\tau \dd\tau'\, C_{\advt,\alpha\beta}^\mu(\tau - \tau')\, \tilde Q^\mu_\alpha(\tau)^\times \tilde Q^\mu_\beta(\tau')^\rarrow \nonumber\\
		&+ \int_0^\tau \dd\tau'\, C_{\rett,\alpha\beta}^\mu(\tau-\tau')\, \tilde Q^\mu_\alpha(\tau)^\times \tilde Q^\mu_\beta(\tau')^\larrow .
\end{align}
Two environments are therefore equivalent as long as the correlation functions
\begin{align}
	C_{\advt,\alpha\beta}^\mu(t) &\equiv \sum\nolimits_n \lambda^\mu_{n\alpha} \lambda^\mu_{n\beta} \tr\bigl[ X^\mu_{n\alpha}(t) X^\mu_{n\beta}\, \rho^\mu_{\eqt,n} \bigr] \quad \text{and} \nonumber\\
	C_{\rett,\alpha\beta}^\mu(t) &\equiv \sum\nolimits_n \lambda^\mu_{n\alpha} \lambda^\mu_{n\beta} \tr\bigl[ X^\mu_{n\beta} X^\mu_{n\alpha}(t)\, \rho^\mu_{\eqt,n} \bigr]
\end{align}
agree for all $\alpha$ and $\beta$.

\subsubsection*{Non-Unitary System Evolution}
Next, we consider adding a non-unitary contribution on the system Hilbert space.
That is, we add a term $\mathbf K(t) \rho(t)$ on the right hand side of Eq.~\eqref{eq:pm_evo}, where $\mathbf K(t)$ is a superoperator acting only on the system space.
This modification has the effect of modifying Eq.~\eqref{eq:app:fv_derivation} into
\begin{equation} \label{eq:app:ref_for_referee}
	\mathbf W(\tau) \equiv \tilde{\mathbf K}(\tau) + \int_0^\tau \dd\tau'\, \expval[\big]{ \tilde{\mathbf L}_\intt(\tau) \tilde{\mathbf L}_\intt(\tau') }_\envt
\end{equation}
%with $\tilde{\mathbf L}_0(t) = \timeordinv\Bigl[ \ee^{-\int_0^t\! \dd\tau\, \mathbf L_\syst(\tau)} \Bigr] \mathbf L_0(t) \timeord\Bigl[ \ee^{\int_0^t\! \dd\tau\, \mathbf L_\syst(\tau)} \Bigr]$.
with $\tilde{\mathbf K}(t)$ the appropriately transformed superoperator.
If $\mathbf K(t)$ is chosen to be
\begin{align}
	\mathbf K(t)\, \rho &= \sum\nolimits_\mu \Gamma^\mu_\advt \bigl( Q^\mu_\phantk(t) \rho Q^\mu_\phantk(t) - Q^\mu_\phantk(t)^2 \rho \bigr) \nonumber\\
		&\quad + \sum\nolimits_\mu \Gamma^\mu_\rett \bigl( Q^\mu_\phantk(t) \rho Q^\mu_\phantk(t) - \rho Q^\mu_\phantk(t)^2 \bigr) , \label{eq:app:terminator}
\end{align}
where $\Gamma^\mu_\advt$ and $\Gamma^\mu_\rett$ are some rates, adding the non-unitary term thus has the same effect as adding $\delta$-terms to the correlation functions:
\begin{align}
	C^\mu_\advt(t) &\to C^\mu_\advt(t) + \Gamma^\mu_\advt \delta(t) , \nonumber\\
	C^\mu_\rett(t) &\to C^\mu_\rett(t) + \Gamma^\mu_\rett \delta(t) .
\end{align}
The term \eqref{eq:app:terminator} is known as an Ishizaki-Tanimura terminator in the HEOM literature \cite{IshizakiJPhysSocJpn2005}.
It may be used to capture short-time features of correlation functions that cannot be fitted well with a multi-exponential ansatz.

\subsubsection*{Modified Interaction Terms}
Here, we consider modifications of the interaction terms.
There is a large number of possible modifications; let us consider for example
\begin{equation} \label{eq:app:gen3}
	\mathbf L_\intt(t)\, \rho \equiv -\ii \sum\nolimits_{\mu n} \lambda^\mu_n\, \bigl( Q^\mu_\phantq(t) Y^\mu_n \rho - \rho Q^\mu_\phantq(t) X^\mu_n \bigl) ,
\end{equation}
where $X^\mu_n$ and $Y^\mu_n$ are unrelated, arbitrary operators.
Plugging this interaction term into Eq.~\eqref{eq:app:fv_derivation}, we find that such environments are equivalent as long as the following four correlation functions agree:
\begin{align}
	C_{(1)}^\mu(t) &\equiv \sum\nolimits_n (\lambda^\mu_n)^2 \tr\bigl[ Y^\mu_n(t) Y^\mu_n\, \rho^\mu_{\eqt,n} \bigr] , \nonumber\\
	C_{(2)}^\mu(t) &\equiv \sum\nolimits_n (\lambda^\mu_n)^2 \tr\bigl[ X^\mu_n X^\mu_n(t)\, \rho^\mu_{\eqt,n} \bigr] , \nonumber\\
	C_{(3)}^\mu(t) &\equiv \sum\nolimits_n (\lambda^\mu_n)^2 \tr\bigl[ X^\mu_n(t) Y^\mu_n\, \rho^\mu_{\eqt,n} \bigr] \text{ and} \nonumber\\
	C_{(4)}^\mu(t) &\equiv \sum\nolimits_n (\lambda^\mu_n)^2 \tr\bigl[ X^\mu_n Y^\mu_n(t)\, \rho^\mu_{\eqt,n} \bigr] .
\end{align}
In this work, we focus on the case $X^\mu_n = Y^\mu_n$ where only two correlation functions need to be matched.
The generalization above, or similar modifications of the interaction terms, might however be able to optimize our results a bit further.

\subsubsection*{Non-Equilibrium Environments}
So far, we have only considered stationary initial states for the unitary and pseudomode environments.
However, this assumption can be relaxed \cite{TamascelliPhysRevLett2018}, as long as the initial states remain Gaussian.
The pseudomode framework can thus also be applied to environments with non-thermal initial states, such as squeezed states.
It has been shown that non-thermal initial states can be used as a thermodynamic resource and thus, for example, increase the performance of quantum heat engines \cite{RossnagelPhysRevLett2014}.

We first review the situation for unitary environments.
If a unitary environment is initially in a non-stationary state $\rho^\mu_0$, its two-time correlation function will generally depend explicitly on both times, instead of just their difference:
\begin{align}
	C^\mu_\phantk(\tau, \tau') &\equiv \tr[ X_\phantk^\mu(\tau) X_\phantk^\mu(\tau') \rho^\mu_0 ] \nonumber\\
		&= \tr[ X_\phantk^\mu(\tau - \tau') X^\mu_\phantk \rho^\mu_0(\tau') ] .
\end{align}
Here, $\rho^\mu_0(\tau') \equiv \exp(-\ii H^\mu_\envt \tau')\, \rho^\mu_0\, \exp(\ii H^\mu_\envt \tau')$ is the time-evolved environment state.
We note that the relation
\begin{equation}
	C^\mu_\phantk(\tau, \tau')^\ast = C^\mu_\phantk(\tau', \tau) = \tr[ X_\phantk^\mu X^\mu_\phantk(\tau - \tau') \rho^\mu_0(\tau') ]
\end{equation}
holds, and therefore $\comm{X^\mu(\tau)}{X^\mu(\tau')} = 2\ii\, \Im[ C^\mu(\tau, \tau') ]$.

We now consider non-equilibrium initial states $\rho^\mu_{0,n}$ for the pseudomodes, and define the correlation functions
\begin{align}
	C^\mu_{\advt,n}(\tau, \tau') &\equiv \tr\bigl[ X^\mu_n(\tau - \tau') X^\mu_n\, \rho^\mu_{0,n}(\tau') \bigr] \quad \text{and} \nonumber\\
	C^\mu_{\rett,n}(\tau, \tau') &\equiv \tr\bigl[ X^\mu_n X^\mu_n(\tau - \tau')\, \rho^\mu_{0,n}(\tau') \bigr] .
\end{align}
in analogy to Eq.~\eqref{eq:app:pm_cfcts}.
Note that $\rho^\mu_{0,n}(\tau') \equiv \exp(\mathbf L^\mu_n \tau') \rho^\mu_0$ are the time-evolved states, and $X^\mu_n(\tau - \tau')$ are still Heisenberg picture operators as defined in Eq.~\eqref{eq:heisenberg}.
Generalizing Eq.~\eqref{eq:app:corr_derivation} to this situation, we find
\begin{align}
	\tr\bigl[ (X^\mu_n)^i_\tau (X^\mu_n)^\rarrow_{\tau'} \rho^\mu_{\eqt,n} \bigr] &= C^\mu_{\advt,n}(\tau, \tau') \quad \text{and} \nonumber\\
	\tr\bigl[ (X^\mu_n)^i_\tau (X^\mu_n)^\larrow_{\tau'} \rho^\mu_{\eqt,n} \bigr] &= C^\mu_{\rett,n}(\tau, \tau')
\end{align}
for any $i \in \{ \larrow, \rarrow \}$.

From here on, the derivation proceeds exactly as before.
We must therefore match the correlation functions as follows:
\begin{align}
	C^\mu_\phantk(\tau, \tau') &= \sum\nolimits_n (\lambda^\mu_n)^2\, C^\mu_{\advt,n}(\tau, \tau') , \nonumber\\
	C^\mu_\phantk(\tau, \tau')^\ast &= \sum\nolimits_n (\lambda^\mu_n)^2\, C^\mu_{\rett,n}(\tau, \tau') .
\end{align}

\subsection{Dissipatons} \label{app:dissipatons}

Until now, we have assumed that the full state $\rho$ is a matrix, that is, an element of the space
\begin{equation}
	\rho \in (\mathcal H_\syst^\phantk \otimes \mathcal H_\envt') \otimes (\mathcal H_\syst^\phantk \otimes \mathcal H_\envt')^\ast ,
\end{equation}
where $\mathcal H_\syst$ is the system Hilbert space, $\mathcal H_\envt'$ the replacement environment Hilbert space and the star denotes the dual.
Removing this assumption is our final generalization to consider.
Instead, we consider a state space of the form
\begin{equation}
	\rho \in \mathcal H_\syst^\phantk \otimes \mathcal H_\syst^\ast \otimes \mathcal H_\disst^\phantk ,
\end{equation}
where $\mathcal H_\disst$ is an arbitrary space.
The partial trace operation must then be replaced with an analogous operation, that is, a linear functional $\varphi: \mathcal H_\disst \to \mathbb C$.

We must still make the basic assumptions that the free evolution on $\mathcal H_\disst$ preserves $\varphi$, that it has a Gaussian stationary state, and that the commutators of the relevant environment coupling operators in an interaction picture are central.
The free evolution preserving $\varphi$ means that $\varphi \circ \mathbf L_\disst = 0$ if $\mathbf L_\disst$ is the generator of the free evolution.
In fact, the first assumption can alternatively be formulated as ``$\mathbf L_\disst$ has a zero eigenvalue'', since the existence of a zero eigenvalue guarantees the existence of a preserved functional $\varphi$, which is the corresponding left eigenvector.
Under these assumptions, the derivation of the influence functional [Eqs.~\eqref{eq:app:inf_functional} and \eqref{eq:app:fv_derivation}] procedes as before.
Note that the operation $\expval{\bullet}_\envt$ appearing in Eq.~\eqref{eq:app:fv_derivation} then refers to a partial application of $\varphi$.

Following this line of thought, one obtains the dissipatons introduced in Refs.~\cite{WangJChemPhys2022, LiArXiv240117255Quant-Ph2024}.
Assume that the time evolution is given by
\begin{equation}
	\partial_t \rho(t) = [ \mathbf L_\syst(t) + \mathbf L_\disst + \mathbf L_\intt(t) ]\, \rho(t) ,
\end{equation}
where $\mathbf L_\disst\, \rho \equiv -\ii \sum_{\mu n} \Omega^\mu_n\, b^{\mu\dagger}_n b^\mu_n\, \rho$ and
\begin{equation}
	\mathbf L_\intt(t)\, \rho \equiv -\ii \sum\nolimits_{\mu n} \bigl[ Q_\phantk^\mu(t) Y^\mu_n \rho(t) - X^\mu_n \rho(t) Q_\phantk^\mu(t) \bigr] .
\end{equation}
Here, $b^{\mu\dagger}_n$ and $b^\mu_n$ are the raising and lowering operators of the dissipatons, and $\Omega^\mu_n$ their (usually complex-valued) frequencies.
If the coupling operators $X^\mu_n$ and $Y^\mu_n$ are linear in $b^{\mu\dagger}_n$ and $b^\mu_n$, our three basic assumptions are all satisfied.
In particular, the stationary state $\ket{0}_\disst$ is Gaussian, and the functional $\varphi \equiv \bra{0}_\disst$ is invariant.

We specify the coupling operators to be
\begin{equation}
	X^\mu_n \equiv b^{\mu\dagger}_n + \lambda^\mu_n b^\mu_n , \quad
	Y^\mu_n \equiv b^{\mu\dagger}_n + \Lambda^\mu_n b^\mu_n
\end{equation}
with some complex constants $\lambda^\mu_n$ and $\Lambda^\mu_n$.
We then obtain
\begin{align}
	&\expval[\big]{ \tilde{\mathbf L}_\intt(\tau) \tilde{\mathbf L}_\intt(\tau') }_\envt
	= \bra{0}_\disst\, \tilde{\mathbf L}_\intt(\tau) \tilde{\mathbf L}_\intt(\tau')\, \ket{0}_\disst \nonumber\\
	&= \sum\nolimits_{\mu n} \bra{0}_\disst\, X^\mu_n(t) X^\mu_n(\tau')\, \ket{0}_\disst\, \tilde Q_\phantk^\mu(\tau)^\times \tilde Q_\phantk^\mu(\tau')^\rarrow \nonumber\\
	&\quad - \sum\nolimits_{\mu n} \bra{0}_\disst\, Y^\mu_n(t) Y^\mu_n(\tau')\, \ket{0}_\disst\, \tilde Q_\phantk^\mu(\tau)^\times \tilde Q_\phantk^\mu(\tau')^\larrow .
\end{align}
Noting that
\begin{align}
	\bra{0}_\disst\, X^\mu_n(t) X^\mu_n(\tau')\, \ket{0}_\disst &= \lambda^\mu_n \exp\bigl[ \ii \Omega^\mu_n (\tau - \tau') \bigr] \quad \text{and} \nonumber\\
	\bra{0}_\disst\, Y^\mu_n(t) Y^\mu_n(\tau')\, \ket{0}_\disst &= \Lambda^\mu_n \exp\bigl[ \ii \Omega^\mu_n (\tau - \tau') \bigr] ,
\end{align}
and comparing with Eq.~\eqref{eq:influence_phase}, we obtain the conditions
\begin{equation}
	C^\mu(t) = \sum\nolimits_n \lambda^\mu_n\, \ee^{\ii \Omega^\mu_n t}
		\quad\text{and}\quad
	C^\mu(t)^\ast = \sum\nolimits_n \Lambda^\mu_n\, \ee^{\ii \Omega^\mu_n t}
\end{equation}
for $t \geq 0$.
The constants $\lambda^\mu_n$, $\Lambda^\mu_n$ and $\Omega^\mu_n$ can be determined from these conditions to find a replacement dissipaton model.

\section{Pseudomode Time Evolution} \label{app:pms}

In this appendix, we will discuss the generator \eqref{eq:pm_lindblad}.
It is obviously a linear, trace-preserving superoperator.
Its unique stationary state is given by Eq.~\eqref{eq:lindblad_ss}; this statement can be proven exactly like in the case of a completely positive Lindblad equation.
The second and third of our assumptions follow from the fact that the superoperators $(X^\mu_n)^i_t$ ($i \in \{\larrow, \rarrow\}$) remain linear combinations of $(b^\mu_n)^\larrow$, $(b^\mu_n)^\rarrow$, $(b^{\mu\dagger}_n)^\larrow$ and $(b^{\mu\dagger}_n)^\rarrow$ at all times.
To see this, note that
\begin{equation} \label{eq:app:pms1}
	(X^\mu_n)^i_t = \ee^{-(\mathbf L^\mu_n)^\times t} (X^\mu_n)^i
\end{equation}
for $i \in \{\larrow, \rarrow\}$, where $(\mathbf L^\mu_n)^\times \bullet \equiv \comm{\mathbf L^\mu_n}{\bullet}$ is a commutator of superoperators.
It thus suffices to show that $(\mathbf L^\mu_n)^\times$ maps the subspace of superoperators spanned by $(b^\mu_n)^\larrow$, $(b^\mu_n)^\rarrow$, $(b^{\mu\dagger}_n)^\larrow$ and $(b^{\mu\dagger}_n)^\rarrow$ to itself.
We calculate the action of $(\mathbf L^\mu_n)^\times$ on each of these basis elements and find that we do not leave the subspace:
\begin{align}
	\mathbf L^\times b^\larrow &= \bigl( \ii\Omega - \Gamma (2N {+} 1) / 2 \bigr)\, b^\larrow + \Gamma (N {+} 1)\, b^\rarrow , \nonumber\\
	\mathbf L^\times b^\rarrow &= \bigl( \ii\Omega + \Gamma (2N {+} 1) / 2 \bigr)\, b^\rarrow - \Gamma N\, b^\larrow , \nonumber\\
	\mathbf L^\times (b^\dagger)^\larrow &= \bigl( -\ii\Omega + \Gamma (2N {+} 1) / 2 \bigr)\, (b^\dagger)^\larrow - \Gamma N\, (b^\dagger)^\rarrow , \nonumber\\
	\mathbf L^\times (b^\dagger)^\rarrow &= \bigl( -\ii\Omega - \Gamma (2N {+} 1) / 2 \bigr)\, (b^\dagger)^\rarrow + \Gamma (N {+} 1)\, (b^\dagger)^\larrow . \label{eq:app:pms2}
\end{align}
We omitted all indices for the sake of presentation.

With the calculation \eqref{eq:app:pms2}, we have gained a matrix representation of $(\mathbf L^\mu_n)^\times$.
Using Eq.~\eqref{eq:app:pms1}, it is thus straightforward to find explicit expressions for $(X^\mu_n)^i_t$.
Plugging these expressions into the identities \eqref{eq:app:corr_derivation}, one obtains Eq.~\eqref{eq:pms_cfcts} after a short calculation.

\section{Quantum Jump Trajectories} \label{app:qjt}

\subsection{Generalized Ansatz}

We will consider a complex scalar $\mu$ which evolves together with the double state $\Psi$ and aim to write the state $\rho$ as the statistical average
\begin{equation}
	\rho = \mathbb E\{ \rho_{\mu,\Psi} \} \equiv \mathbb E\{ \mu\, \ketbra{\psi_1}{\psi_2} \} .
\end{equation}
The scalar and the double state follow the following coupled stochastic differential equations:
\begin{align}
	\dd \ket{\psi_1} &= \Bigl[ -\ii H + \frac 1 2 f_1 - \frac 1 2 \sum\nolimits_\alpha \gamma_\alpha L_\alpha^\dagger L_\alpha \Bigr] \ket{\psi_1}\, \dd t \nonumber\\
		&\quad + \sum\nolimits_\alpha \bigl[ g_1^\alpha\, L_\alpha \ket{\psi_1} - \ket{\psi_1} \bigr]\, \dd N_\alpha , \nonumber\displaybreak[0]\\
	\dd \ket{\psi_2} &= \Bigl[ -\ii H^\dagger + \frac 1 2 f_2^\ast - \frac 1 2 \sum\nolimits_\alpha \gamma_\alpha^\ast L_\alpha^\dagger L_\alpha \Bigr] \ket{\psi_2}\, \dd t \nonumber\\
		&\quad + \sum\nolimits_\alpha \bigl[ g_2^{\alpha\ast}\, L_\alpha \ket{\psi_2} - \ket{\psi_2} \bigr]\, \dd N_\alpha \quad \text{and} \nonumber\displaybreak[0]\\
	\dd \mu &= f_\mu\, \mu\, \dd t + \sum\nolimits_\alpha \bigl[ g_\mu^\alpha - 1 \bigr]\, \mu\, \dd N_\alpha . \label{eq:app:sde}
\end{align}
Again, $\dd N_\alpha$ are differentials of independent Poisson processes with conditional expectation values
\begin{equation}
	\mathbb E\{ \dd N_\alpha \mid \mu, \Psi \} = r_\alpha\, \dd t .
\end{equation}

Like the rates $r_\alpha > 0$, the newly introduced degrees of freedom $f_i$ and $g_i^\alpha$ are \emph{a priori} undetermined functions of $\mu$ and $\Psi$.
To make $\rho$ satisfy the master equation \eqref{eq:general_lindblad}, they must satisfy the following relations:
\begin{equation} \label{eq:app:lindblad_cond}
	f_\mu + \frac{f_1 + f_2}{2} = \sum\nolimits_\alpha r_\alpha
		\quad \text{and} \quad
	r_\alpha\, g_\mu^\alpha g_1^\alpha g_2^\alpha = \gamma_\alpha .
\end{equation}
The functions $g_i^\alpha$ have the effect of rescaling the quantities $\ket{\psi_1}$, $\ket{\psi_2}$ and $\mu$ after a jump in the corresponding dissipative channel such that their product $\rho_{\mu,\Psi}$ remains the same.
Similarly, the functions $f_i^\alpha$ redistribute weight between these quantities during the continuous part of their evolution.

For a given realization of the random variables $\dd N_\alpha$, the resulting state $\rho_{\mu,\Psi}$ here will thus be identical to the state $\rho_\Psi$ in the main text.
Hence, the generalization considered here does not remove the issue of exponential growth found there.
We can however try applying various choices of the new degrees of freedom in order to minimize the impact of this issue, or simply to make the scheme easier to implement numerically.
In the following sections, we will explore some of these choices.
We will frequently use the notation $\rho_\Psi \equiv \ketbra{\psi_1}{\psi_2}$, $\expval{\bullet}_\Psi \equiv \bra{\psi_2} \bullet \ket{\psi_1}$, $\norm{\psi_i}^2 \equiv \braket{\psi_i}{\psi_i}$ and $\expval{\bullet}_i \equiv \bra{\psi_i} \bullet \ket{\psi_i}$ for $i = 1,2$.

\subsection{Constant Scalar}

We first take a step back and consider the scheme without the scalar, setting $f_\mu = 0$ and $g_\mu^\alpha = 1$.
We thus consider the evolution of a double state $\rho_\Psi \equiv \ketbra{\psi_1}{\psi_2}$ as in the main text, but with additional degrees of freedom $f_i$ and $g_i^\alpha$.
Following Eq.~\eqref{eq:app:lindblad_cond}, they must satisfy
\begin{equation}
	\frac{f_1 + f_2}{2} = \sum\nolimits_\alpha r_\alpha
		\quad \text{and} \quad
	r_\alpha g_1^\alpha g_2^\alpha = \gamma_\alpha .
\end{equation}
The unraveling shown in the main text is the special case where $f_1 = f_2 = \sum_\alpha r_\alpha$ and $g_1^\alpha = g_2^\alpha = (\gamma_\alpha / r_\alpha)^{1/2}$.

In the main text, we focused on minimizing the fluctuations of $\tr \rho_\Psi$.
We therefore chose the rates $r_\alpha$ such that $\mathbb E\{ \dd( \abs{\tr \rho_\Psi}^2 ) \mid \Psi \}$ is minimal.
Since $\rho_\Psi$ does not depend on the choice of $f_i$ and $g_i^\alpha$, the rates that achieve this goal are still
\begin{equation} \label{eq:app:rate1}
	r_\alpha^\ast = \abs*{ \frac{\gamma_\alpha^\phantq\, \expval{ L_\alpha^\dagger L_\alpha^\phantq }_\Psi}{\tr \rho_\Psi} } .
\end{equation}
as in Eq.~\eqref{eq:mc:rates}, independent of $f_i$ and $g_i^\alpha$.

Alternatively, one could try to minimize, for example, the product of the individual norms.
Computing
\begin{align}
	&\mathbb E\{ \dd( \norm{\psi_1}^2 \norm{\psi_2}^2 ) \mid \Psi \} \nonumber\\
	&= \sum\nolimits_\alpha \Bigl[ r_\alpha\, \norm{\psi_1}^2 \norm{\psi_2}^2 + \frac{\abs{\gamma_\alpha}^2}{r_\alpha} \expval{ L_\alpha^\dagger L_\alpha^\phantq }_1 \expval{ L_\alpha^\dagger L_\alpha^\phantq }_2 \Bigr]\, \dd t \nonumber\\
	&\quad + [ \text{terms not depending on $r_\alpha$, $f_i$ or $g_i^\alpha$} ] ,
\end{align}
we find that this goal is achieved by the rates
\begin{equation} \label{eq:app:rate2}
	r_\alpha^{\text{alt}} = \abs{\gamma_\alpha}\, \sqrt{ \frac{\expval{ L_\alpha^\dagger L_\alpha^\phantq }_1}{\norm{\psi_1}^2} \frac{\expval{ L_\alpha^\dagger L_\alpha^\phantq }_2}{\norm{\psi_2}^2} } ,
\end{equation}
again independent of $f_i$ and $g_i^\alpha$.

Whether we choose the rates according to Eq.~\eqref{eq:app:rate1}, \eqref{eq:app:rate2} or otherwise, we are still free to choose $f_i$ and $g_i^\alpha$.
One possible approach would be to demand $\norm{\psi_1} = \norm{\psi_2}$ on each trajectory at all times.
To this end, we would choose
\begin{align}
	f_i &= \sum\nolimits_\alpha \Bigl[ r_\alpha \mp \Re(\gamma_\alpha) \frac{\expval{ L_\alpha^\dagger L_\alpha^\phantq }_1 - \expval{ L_\alpha^\dagger L_\alpha^\phantq }_2}{\norm{\psi_1}^2 + \norm{\psi_2}^2} \Bigr] \nonumber\\
		&\quad \mp \ii \frac{\expval{H - H^\dagger}_1 + \expval{H - H^\dagger}_2}{\norm{\psi_1}^2 + \norm{\psi_2}^2}
		\quad \text{and} \nonumber\displaybreak[0]\\
	g^\alpha_{1(2)} &= \sqrt{ \frac{\gamma_\alpha^\phantq}{r_\alpha^\phantq} \biggl( \frac{\expval{ L_\alpha^\dagger L_\alpha^\phantq }_{2(1)}}{\expval{ L_\alpha^\dagger L_\alpha^\phantq }_{1(2)}} \biggr)^{1/2} } . \label{eq:app:samenorms}
\end{align}

Let us also briefly consider the expected change of the individual norms, $\delta_i(\Psi) \equiv \mathbb E\{ \dd \norm{\psi_i}^2 \mid \Psi \}$.
Whether these quantities can be both set to zero using $f_i$ and $g_i^\alpha$ depends on $\Psi$ and on the anti-Hermitian part of the Hamiltonian, $H_I \equiv \frac{1}{2\ii} (H - H^\dagger)$.
A detailed analysis shows that $\delta_1(\Psi) = \delta_2(\Psi) = 0$ is possible if and only if the inequality
\begin{align}
	\frac{\expval{H_I}_1}{\norm{\psi_1}^2} - \frac{\expval{H_I}_2}{\norm{\psi_2}^2} \leq -\sum\nolimits_\alpha \abs{\gamma_\alpha}\, \sqrt{ \frac{\expval{ L_\alpha^\dagger L_\alpha^\phantq }_1}{\norm{\psi_1}^2} \frac{\expval{ L_\alpha^\dagger L_\alpha^\phantq }_2}{\norm{\psi_2}^2} }
\end{align}
is satisfied.
Due to this complication, we will not explore this avenue further.
The sum $\delta_1(\Psi) + \delta_2(\Psi)$ relates to the change of the squared norm $\norm{\Psi}^2 = \norm{\psi_1}^2 + \norm{\psi_2}^2$ of the double state.
It can always be set to zero, or indeed to any other value, and is therefore less helpful in guiding us towards practical choices of the rates.
However, we will consider an unraveling with $\delta_1(\Psi) + \delta_2(\Psi) = 0$ in Sec.~\ref{subsec:app:assoc_cptp}.

\subsection{Martingale Scalar} \label{app:martingale}

The additional complex scalar introduced in Eq.~\eqref{eq:app:sde} can be used to absorb the changes of $\tr \rho_\Psi$, i.e., keep $\tr \rho_\Psi = 1$ throughout every trajectory, if we set
\begin{align}
	f_\mu &= \sum\nolimits_\alpha \bigl( r_\alpha^\phantq - \gamma_\alpha^\phantq \expval{ L_\alpha^\dagger L_\alpha^\phantq }_\Psi \bigr) \quad \text{and} \nonumber\\
	g_\mu^\alpha &= \frac{\gamma_\alpha}{r_\alpha} \expval{ L_\alpha^\dagger L_\alpha^\phantq }_\Psi . \label{eq:app:martingale_choice}
\end{align}
Then, the equation of motion for the scalar becomes
\begin{equation}
	\dd\mu = \mu \sum\nolimits_\alpha \Bigl( 1 - \frac{\gamma_\alpha}{r_\alpha} \expval{ L_\alpha^\dagger L_\alpha^\phantq }_\Psi \Bigr) \bigl( r_\alpha\, \dd t - \dd N_\alpha \bigr) .
\end{equation}
We immediately see that the scalar now satisfies the martingale property
\begin{equation}
	\mathbb E\{ \dd\mu \mid \mu, \Psi \} = 0 ,
\end{equation}
which corresponds to the original master equation being trace-preserving.
The fluctuations of this martingale are related to the fluctuations of $\tr \rho_\Psi$ for constant $\mu$,
\begin{equation}
	\mathbb E\{ \dd \abs{\mu}^2 \mid \mu, \Psi \} = \sum\nolimits_\alpha \frac{\abs{\mu}^2}{r_\alpha} \abs*{r_\alpha - \frac{ \mu \gamma_\alpha^\phantq\, \expval{ L_\alpha^\dagger L_\alpha^\phantq }_\Psi }{\mu} }^2\, \dd t ,
\end{equation}
which can be compared to Eq.~\eqref{eq:mc:fluctuations}.

In the case where $H$ is Hermitian and $\gamma_\alpha$ real (but possibly negative), we can choose $f_1 = f_2$ and $g_1^\alpha = g_2^\alpha$ to make the states $\ket{\psi_1}$ and $\ket{\psi_2}$ identical.
In this case, the stochastic differential equation \eqref{eq:app:sde} reduces to the unraveling proposed in Ref.~\cite{DonvilNatCommun2022} where $\mu$ was called the influence martingale.
Note that our and their definition of the rates $r_\alpha$ differ by the factor $\expval{ L_\alpha^\dagger L_\alpha^\phantq }_\Psi$.

If $H$ is non-Hermitian or $\gamma_\alpha$ complex, $\ket{\psi_1}$ and $\ket{\psi_2}$ generally cannot be identical.
However, we can again keep $\norm{\psi_1} = \norm{\psi_2}$ with the choices
\begin{align}
	f_i &= \sum\nolimits_\alpha \Bigl[ \gamma_\alpha^\phantq \expval{ L_\alpha^\dagger L_\alpha^\phantq }_\Psi \mp \Re(\gamma_\alpha^\phantq) \frac{\expval{ L_\alpha^\dagger L_\alpha^\phantq }_1 - \expval{ L_\alpha^\dagger L_\alpha^\phantq }_2}{\norm{\psi_1}^2 + \norm{\psi_2}^2} \Bigr] \nonumber\\
		&\quad \mp \ii \frac{\expval{H - H^\dagger}_1 + \expval{H - H^\dagger}_2}{\norm{\psi_1}^2 + \norm{\psi_2}^2}
		\quad \text{and} \nonumber\displaybreak[0]\\
	g^\alpha_{1(2)} &= \sqrt{ \frac{1}{\expval{ L_\alpha^\dagger L_\alpha^\phantq }_\Psi} \biggl( \frac{\expval{ L_\alpha^\dagger L_\alpha^\phantq }_{2(1)}}{\expval{ L_\alpha^\dagger L_\alpha^\phantq }_{1(2)}} \biggr)^{1/2} } .
\end{align}

We finally remark that $\mu$ being a martingale is due to Eq.~\eqref{eq:app:martingale_choice} and not true in general.
For example, one could determine $f_i$ and $g_i^\alpha$ from the condition $\dd \norm{\psi_1}^2 = \dd \norm{\psi_2}^2 = 0$.
In that case, it is easy to check that $\mathbb E\{ \dd\mu \mid \mu, \Psi \} \neq 0$ in general.

\subsection{CPTP Evolution on the Double Space} \label{subsec:app:assoc_cptp}

In this section, we will show that it is possible to generate an unraveling for the pseudo-Lindblad equation \eqref{eq:general_lindblad} as follows.
First, one generates trajectories for an associated Lindblad equation on the double space.
Second, one computes a complex scalar value $\mu$ for each trajectory.
The ensemble average of $\rho_{\mu,\Psi} \equiv \mu\, \ketbra{\psi_1}{\psi_2}$ (where $\Psi \equiv (\ket{\psi_1}, \ket{\psi_2})$ is the double state) then satisfies the pseudo-Lindblad equation.
Since the associated Lindblad equation is completely positive and trace preserving, and since the trajectories do not depend on $\mu$, the first step can be done using any existing quantum Monte Carlo tool such as QuTiP's \inlinecode{mcsolve} \cite{JohanssonComputPhysCommun2012, JohanssonComputPhysCommun2013}.
This procedure is more convenient than implementing the stochastic process manually, but it requires a specific choice of $r_\alpha$ that might be worse for convergence.

The following is an extension of the technique introduced in Ref.~\cite{DonvilNewJPhys2023} (which QuTiP's \inlinecode{nm\_mcsolve} function is based on), see also Ref.~\cite{HushPhysRevA2015}.
For notational convenience, we will assume that the dissipation channels are labelled by an integer $\alpha \in \{1 \dots n\}$.
We call the Hermitian and anti-Hermitian parts of the Hamiltonian $H_R \equiv \frac 1 2 (H + H^\dagger)$ and $H_I \equiv \frac 1 {2\ii} (H - H^\dagger)$, respectively, and the real and imaginary parts of $\gamma_\alpha$ are $\gamma_{R\alpha}$ and $\gamma_{I\alpha}$.
Define
\begin{align}
	\hat L_\alpha &\equiv \begin{pmatrix} L_\alpha & 0 \\ 0 & L_\alpha \end{pmatrix} \quad \text{($1 \leq \alpha \leq n$)} \quad \text{and} \nonumber\\
	\hat H &\equiv \begin{pmatrix} H_R + \sum_{\alpha=1}^n \frac{\gamma_{I\alpha}}{2} L_\alpha^\dagger L_\alpha^\phantq & 0 \\ 0 & H_R - \sum_{\alpha=1}^n \frac{\gamma_{I\alpha}}{2} L_\alpha^\dagger L_\alpha^\phantq \end{pmatrix} .
\end{align}
Hats denote operators on the double Hilbert space.

We can now construct the associated Lindblad equation.
It will have $(n+1)$ dissipation channels with corresponding rates $\Gamma_\alpha > 0$ ($0 \leq \alpha \leq n$) that can be chosen freely.
To determine the extra Lindblad operator $\hat L_0$, consider the Hermitian operator
\begin{equation}
	\hat X \equiv \sum_{\alpha=1}^n (\Gamma_\alpha - \gamma_{R\alpha}) \hat L_\alpha^\dagger \hat L_\alpha^\phantq + 2 \begin{pmatrix} H_I & 0 \\ 0 & -H_I \end{pmatrix} .
\end{equation}
Assuming that it is bounded, we can find a $\Lambda \in \mathbb R$ such that $\Lambda - \hat X \geq 0$.
We can then find an $\hat L_0 = \operatorname{diag}( L_0^{(1)}, L_0^{(2)} )$ such that
\begin{equation} \label{eq:app:completeness}
	\Gamma_0^\phantq\, \hat L_0^\dagger \hat L_0^\phantq = \Lambda - \hat X .
\end{equation}

The associated, completely positive Lindblad equation for the double state $\hat\rho$ is
\begin{equation} \label{eq:app:assoc_me}
	\partial_t \hat\rho = -\ii \comm{\hat H}{\hat\rho} + \sum_{\alpha=0}^n \Gamma_\alpha^\phantq \bigl( \hat L_\alpha^\phantq \hat\rho \hat L_\alpha^\dagger - \acomm{\hat L_\alpha^\dagger \hat L_\alpha^\phantq}{\hat\rho} / 2 \bigr) .
\end{equation}
The standard unraveling of this master equation into trajectories with $\hat\rho = \mathbb E\{ \Psi \Psi^\dagger \}$ and $\Psi^\dagger \Psi = 1$ is
\begin{align}
	\dd\Psi &= -\ii \hat H \Psi\, \dd t - \frac 1 2 \sum_{\alpha=0}^n \bigl( \Gamma_\alpha^\phantq \hat L_\alpha^\dagger \hat L_\alpha^\phantq - r_\alpha^\phantq \bigr) \Psi\, \dd t \nonumber\\
		&\quad + \sum_{\alpha=0}^n \bigl( \sqrt{\Gamma_\alpha / r_\alpha}\, \hat L_\alpha \Psi - \Psi \bigr)\, \dd N_\alpha ,
\end{align}
where $\dd N_\alpha$ are differentials of independent Poisson processes with
\begin{equation}
	\mathbb E\{ \dd N_\alpha \mid \Psi \} = r_\alpha^{\text{CP2}}\, \dd t , \quad
	r_\alpha^{\text{CP2}} = \Gamma_\alpha^\phantq\, \Psi^\dagger \hat L_\alpha^\dagger \hat L_\alpha^\phantq \Psi .
\end{equation}
This choice of jump rates is necessary to fix $\Psi^\dagger \Psi = 1$ on the trajectory level.

We rewrite this stochastic differential equation in terms of its components and use Eq.~\eqref{eq:app:completeness} to obtain
\begin{align}
	\dd \ket{\psi_1} &= \Biggl[ -\ii H - \frac{\Lambda}{2} - \sum_{\alpha=0}^n \frac{r_\alpha}{2} - \frac 1 2 \sum_{\alpha=1}^n \gamma_\alpha^\phantq L_\alpha^\dagger L_\alpha^\phantq \Biggr] \ket{\psi_1}\, \dd t \nonumber\\
		&\quad + \sum_{\alpha=1}^n \bigl[ \sqrt{(\Gamma_\alpha / r_\alpha)}\, L_\alpha \ket{\psi_1} - \ket{\psi_1} \bigr]\, \dd N_\alpha \nonumber\\
		&\quad + \bigl[ \sqrt{(\Gamma_0 / r_0)}\, L_0^{(1)} \ket{\psi_1} - \ket{\psi_1} \bigr]\, \dd N_0 \quad \text{and} \nonumber\displaybreak[0]\\
	\dd \ket{\psi_2} &= \Biggl[ -\ii H^\dagger - \frac{\Lambda}{2} - \sum_{\alpha=0}^n \frac{r_\alpha}{2} - \frac 1 2 \sum_{\alpha=1}^n \gamma_\alpha^{\ast\phantq} L_\alpha^\dagger L_\alpha^\phantq \Biggr] \ket{\psi_2}\, \dd t \nonumber\\
		&\quad + \sum_{\alpha=1}^n \bigl[ \sqrt{(\Gamma_\alpha / r_\alpha)}\, L_\alpha \ket{\psi_2} - \ket{\psi_2} \bigr]\, \dd N_\alpha \nonumber\\
		&\quad + \bigl[ \sqrt{(\Gamma_0 / r_0)}\, L_0^{(2)} \ket{\psi_2} - \ket{\psi_2} \bigr]\, \dd N_0 .
\end{align}
Aside from the $\dd N_0$-terms, this equation has the same shape as Eq.~\eqref{eq:app:sde}.
In analogy to there, we can make $\rho = \mathbb E\{ \rho_{\mu,\Psi} \}$ satisfy the pseudo-Lindblad equation by introducing a scalar $\mu$ with
\begin{equation}
	\dd\mu = \Lambda \mu\, \dd t + \sum_{\alpha=1}^n \Bigl[ \frac{\gamma_\alpha}{\Gamma_\alpha} - 1 \Bigr] \mu\, \dd N_\alpha - \mu\, \dd N_0 .
\end{equation}
Equation \eqref{eq:app:assoc_me} can be simulated without knowledge of $\mu$ and the value of $\mu$ determined afterwards.
Given a trajectory with jump counts $N_\alpha$ in the respective dissipation channels, the value is $\mu = 0$ if $N_0 \geq 1$ and
\begin{equation} \label{eq:martingale_exponential_growth}
	\mu = \ee^{\Lambda t} \prod_{\alpha=1}^n \Bigl( \frac{\gamma_\alpha}{\Gamma_\alpha} \Bigr)^{N_\alpha}
\end{equation}
otherwise.

\subsection{Comparison of Unravelings} \label{app:stability} \label{app:mc:comparison}

\begin{figure}
	\centering
	\includegraphics[scale=1]{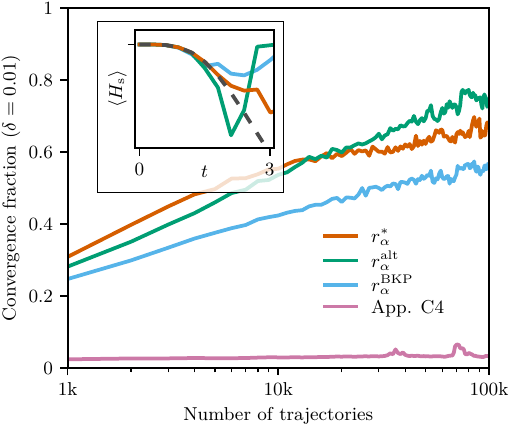}
	\caption{
		Comparisons of different unravelings for Example 1.
		The convergence fraction, shown on the y-axis, is the fraction of the total time interval $t \in [0, 75]$ on which the Monte Carlo estimate of $\expval{H_\syst}$ differs from the exact result by at most $\delta$.
		To produce the plot, we generated a total of $N_{\text{traj}} \approx 500\textrm{k}$ trajectories for each unraveling.
		Given a number $N$ on the x-axis, we grouped these trajectories into $\operatorname{floor}(N_{\text{traj}} / N)$ bunches of $N$ trajectories and calculated the convergence fraction for each bunch.
		The data shown here is the average over all bunches.
		The inset gives more details for the unraveling from Appendix \ref{subsec:app:assoc_cptp}. %, where the convergence fraction never exceeds $7\%$.
		It is an analogue of the upper panel of Fig.~\ref{fig:ex1:trajectories}, using this alternative unraveling and the much shorter time interval $t \in [0,3]$.
		The blue, green and red curves are averages of 1k, 10k and 500k trajectories, respectively.
	}
	\label{fig:ex1:convergence}
\end{figure}

In Fig.~\ref{fig:ex1:convergence}, we compare some of the unravelings introduced so far.
Specifically, we consider the unraveling \eqref{eq:sde} with the rates $r_\alpha^\ast$ defined in \eqref{eq:mc:rates} and used in the main text, with the rates $r_\alpha^{\text{BKP}}$ defined in \eqref{eq:mc:breuer} and with the alternative rates $r_\alpha^{\text{alt}}$ defined in \eqref{eq:app:rate2}, as well as the unraveling discussed in Appendix \ref{subsec:app:assoc_cptp} (with the choice $\Gamma_\alpha = \abs{\gamma_\alpha}$).
We find that the rates $r_\alpha^\ast$ and $r_\alpha^{\text{alt}}$ perform similarly well, and outperform the rates $r_\alpha^{\text{BKP}}$ of Ref.~\cite{BreuerPhysRevA1999}.
The fact that $r_\alpha^\ast$ is beaten by $r_\alpha^{\text{alt}}$ at large numbers of trajectories might be an artifact of insufficient sample size.
All curves appear to grow approximately logarithmically with the number of trajectories, confirming the exponentially growing instability discussed in Sec.~\ref{subsec:qjt}.

The unraveling introduced in Appendix \ref{subsec:app:assoc_cptp} performs very badly in this example.
It shows no convergence for $t \gtrsim 2$ even when averaging over 500k trajectories.
This result can be understood by considering the growth of the scalar component following Eq.~\eqref{eq:martingale_exponential_growth}.
Since its typical size is $\mu \sim \ee^{\Lambda t}$, we expect the number of remaining trajectories with non-zero $\mu$ to scale with $\ee^{-\Lambda t}$.
For our Example 1, we find $\Lambda \approx 3.902$.
With $N=500\textrm{k}$ initial trajectories, the expected number of remaining trajectories becomes less than one at $t = \log(N) / \Lambda \approx 3.363$.

\section{Inversion Sampling of Waiting Times} \label{app:gillespie}

Consider a random variable $X$ undergoing a general piecewise deterministic jump process of the form
\begin{equation}
	\dd X = \mathcal L_t X\, \dd t + \sum\nolimits_\alpha (\mathcal J_\alpha X - X)\, \dd N_\alpha .
\end{equation}
Here, $\dd N_\alpha$ are increments of independent Poisson processes with $\mathbb E\{\dd N_\alpha \mid X\} = r_\alpha[X]\, \dd t$ and we assume that the jump channels are labelled by $\alpha \in \{1 \dots n\}$.
The functions $\mathcal L_t$ and $J_\alpha$ are not required to be linear.
In this section, we discuss how to algorithmically generate trajectories according to this stochastic differential equation.
We consider trajectories on the time interval $t_0 \leq t \leq t_\fint$ with the initial condition $X(t=t_0) = X_0$.

The equation immediately invites the following algorithmic interpretation.\\
\begin{algorithm}[H]
	\caption{Naive Implementation \label{algo:1}}
	\DontPrintSemicolon
	\SetAlgoNoEnd
	\SetAlgoNoLine
	
	Initialize $t \gets t_0$, $X \gets X_0$.\;
	\While{$t < t_\fint$}{
		Generate a random integer $a$ ($0 \leq a \leq n$) according to the probabilities $p_a = r_a[X]\, \dd t$ for $a \geq 1$ and $p_0 = 1 - \sum_\alpha r_\alpha[X]\, \dd t$.\;
		\If{$a=0$}{
			Update $t \gets t + \dd t$ and $X \gets X + \mathcal L_t X\, \dd t$.\;
		}
		\Else{
			Update $t \gets t + \dd t$ and $X \gets \mathcal J_a X$.\;
		}
	}
\end{algorithm}
\noindent
In practice, the time step $\dd t$ must be chosen finite and small enough.
We imagine an idealized algorithm where $\dd t$ is infinitesimal ($\dd t^2 = 0$).

A jump record $R$ is the combined information about the number $N \geq 0$ of jumps on a trajectory, the jump channels $\alpha_k$ for $1 \leq k \leq N$ and the corresponding jump times $t_k$ (with $t_{k+1} > t_k$).
If the algorithm above generates a certain jump record $R$, the corresponding jump trajectory is
\begin{equation} \label{eq:app:general_trajectory}
	X_R(t) = \mathcal U(t, t_K) \circ \mathcal J_{\alpha_k} \circ \cdots \circ \mathcal J_{\alpha_1} \circ \mathcal U(t_1, t_0) X_0 .
\end{equation}
Here, $K$ is the largest index with $t_K < t$, $\circ$ denotes function composition and $\mathcal U$ the solution of the deterministic evolution.
That is, $\mathcal U(t, t_k)X$ is the solution of the initial value problem $\partial_t U(t, t_k)X = \mathcal L_t \circ U(t, t_k)X$ and $\mathcal U(t_k, t_k)X = X$.
We read off the probability of the jump record from the algorithm:
\begin{align}
	P[R]\, \dd \vec t &= \exp\biggl[ -\int_{t_0}^{t_\fint} \sum\nolimits_\alpha r_\alpha[X_R(\tau)]\, \dd \tau \biggr] \nonumber\\
	&\quad \times \prod_{k=1}^N r_{\alpha_k}[X_R(t_k)]\, \dd t_k ,
\end{align}
where we used that $1 - \sum_\alpha r_\alpha[X]\, \dd t = \ee^{-\sum_\alpha r_\alpha[X]\, \dd t}$ and set $\dd\vec t \equiv \dd t_1 \cdots \dd t_N$.

With this algorithm, the deterministic part of the evolution can only be integrated a small step $\dd t$ at a time even if jumps are rare.
To find a more efficient approach, consider the state after the $k$-th jump, $X_k \equiv \mathcal J_{\alpha_k} X_R(t_k)$.
The \emph{a priori} probability of finding no other jump until the time $t$ is given by
\begin{equation}
	P_0^{(k)}(t) \equiv \ee^{-\int_{t_k}^t \sum_\alpha r_\alpha[\mathcal U(\tau, t_k) X_k]\, \dd\tau}
\end{equation}
and the \emph{a priori} distribution of waiting times until the next jump is $\mathcal W^{(k)}(t) = -\partial_t P_0^{(k)}(t)$.
The time of the next jump can be determined directly by applying inversion sampling of the distribution $\mathcal W^{(k)}$.
With this approach, $t_{k+1}$ is determined by the condition $P_0^{(k)}(t_{k+1}) = \lambda$, where $\lambda \in [0, 1]$ is chosen uniformly.
We thus arrive at the algorithm below \cite{GillespieAnnuRevPhysChem2007}.
Further improvements of this algorithm are discussed in Ref.~\cite{RadaelliArXiv230315405Quant-Ph2023}, but were not included in the simulations performed for this paper.
\\
\begin{algorithm}[H]
	\caption{Gillespie \label{algo:2}}
	\DontPrintSemicolon
	\SetAlgoNoEnd
	\SetAlgoNoLine
	\SetKwFor{Repeat}{Repeat}{}{EndRepeat}
	
	\setlength\abovedisplayskip{0pt}
	\setlength\belowdisplayskip{0pt}
	
	Initialize $t_k \gets t_0$, $X_k \gets X_0$.\;
	\Repeat{}{
		Generate $\lambda \in [0, 1]$ uniformly.\;
		Integrate
			\begin{equation*} \textstyle \partial_t X = \mathcal L_t X \text{ and } \partial_t P_0 = -P_0 \sum_\alpha r_\alpha[X] \end{equation*}
		with the initial conditions
			\begin{equation*} \textstyle X(t_k) = X_k \text{ and } P_0(t_k) = 1 \end{equation*}
		until one of the these conditions is reached:\;
		\If{$t = t_\fint$}{
			\Return $X$.\;
		}
		\If{$P_0 = \lambda$}{
			Generate a random integer $\alpha$ ($1 \leq \alpha \leq n$) according to the weights $r_\alpha[X]$.\;
			Update $t_k \gets t$ and $X_k \gets \mathcal J_{\alpha} X$.\;
		}
	}
\end{algorithm}
\mbox{} % layout fix

Clearly, both algorithms generate the same trajectories $X_R(t)$ described in Eq.~\eqref{eq:app:general_trajectory}.
The second algorithm generates a jump record with the probability
\begin{align}
	P'[R]\, \dd\vec t &= \operatorname{Prob}[ \lambda_{N+1} \leq P_0^{(N)}(t_\fint) ] \nonumber\\
		&\quad \times \prod_{k=1}^N \operatorname{Prob}[ P_0^{k-1}(t_k) < \lambda_k \leq P_0^{k-1}(t_k - \dd t_k) ] \nonumber\\
		&\quad \times \prod_{k=1}^N \frac{r_{\alpha_k}[X_R(t_k)]}{\sum_\beta r_\beta[X_R(t_k)]} .
\end{align}
Using the identities $\operatorname{Prob}[ a < \lambda \leq b ] = b - a$ and
\begin{equation}
	P_0^{k-1}(t_k - \dd t_k) = P_0^{k-1}(t_k) \Bigl( 1 + \sum\nolimits_\beta r_\beta[X_R(t_k)]\, \dd t_k \Bigr) ,
\end{equation}
we find that $P'[R] = P[R]$.
The two algorithms are therefore equivalent.
%\vspace*{12em} % layout fix

\vbadness=10000
\hbadness=10000

\end{document}